\makeatother

\documentclass[twocolumn]{aastex62}
\usepackage[]{graphicx,tabularx,ctable,color,colortbl,amsmath,float}
\usepackage[T1]{fontenc}
\definecolor{Gray}{gray}{0.9}

\shortauthors{}

\begin{document}

\title{\textbf{Solar Radio Spikes and Type IIIb Striae Manifestations of Sub-second Electron Acceleration Triggered by a Coronal Mass Ejection}}

\author[0000-0003-1967-5078]{Daniel L. Clarkson}
\affiliation{School of Physics \& Astronomy, University of Glasgow, Glasgow, G12 8QQ, UK}

\author[0000-0002-8078-0902]{Eduard P. Kontar}
\affiliation{School of Physics \& Astronomy, University of Glasgow, Glasgow, G12 8QQ, UK}

\author[0000-0002-6872-3630]{Nicole Vilmer}
\affiliation{LESIA, Observatoire de Paris, Universit\'{e} PSL, CNRS, Sorbonne Universit\'{e}, Universit\'{e} de Paris, 5 place Jules Janssen, 92195 Meudon, France}
\affiliation{Station de Radioastronomie de Nan\c{c}ay, Observatoire de Paris, CNRS, PSL, Universit\'{e} d'Orl\'{e}ans, Nan\c{c}ay, France}

\author[0000-0003-2291-4922]{Mykola Gordovskyy}
\affiliation{Department of Physics \& Astronomy, University of Manchester, Manchester M13 9PL, UK}

\author[0000-0002-1810-6706]{Xingyao Chen}
\affiliation{School of Physics \& Astronomy, University of Glasgow, Glasgow, G12 8QQ, UK}

\author[0000-0002-4389-5540]{Nicolina Chrysaphi}
\affiliation{LESIA, Observatoire de Paris, Universit\'{e} PSL, CNRS, Sorbonne Universit\'{e}, Universit\'{e} de Paris, 5 place Jules Janssen, 92195 Meudon, France}
\affiliation{School of Physics \& Astronomy, University of Glasgow, Glasgow, G12 8QQ, UK}

\begin{abstract}
    Understanding electron acceleration associated with magnetic energy release at sub-second scales presents a major challenges in solar physics. Solar radio spikes observed as sub-second, narrow bandwidth bursts with $\Delta{f}/f\sim10^{-3}-10^{-2}$ are indicative of sub-second evolution of the electron distribution. We present a statistical analysis of frequency, and time-resolved imaging of individual spikes and Type IIIb striae associated with a coronal mass ejection (CME). LOFAR imaging reveals that co-temporal ($<2$~s) spike and striae intensity contours almost completely overlap. On average, both burst types have similar source size with fast expansion at millisecond scales. The radio source centroid velocities are often superluminal, and independent of frequency over $30-45$~MHz. The CME perturbs the field geometry, leading to increased spike emission likely due to frequent magnetic reconnection. As the field restores towards the prior configuration, the observed sky-plane emission locations drift to increased heights over tens of minutes. Combined with previous observations above $1$~GHz, average decay time and source size estimates follow $\sim1/f$ dependency over three decades in frequency, similar to radio-wave scattering predictions. Both time and spatial characteristics of the bursts between $30-70$~MHz are consistent with radio-wave scattering with strong anisotropy of the density fluctuation spectrum. Consequently, the site of radio-wave emission does not correspond to the observed burst locations and implies acceleration and emission near the CME flank. The bandwidths suggest intrinsic emission source sizes $<1$~arcsec at $30$~MHz, and magnetic field strengths a factor of two larger than average in events that produce decameter spikes.
\end{abstract}


\section{Introduction}

Radio bursts are routinely emitted in the outer solar corona due to the acceleration of energetic electrons in solar flares and coronal mass ejections (CMEs). Among many solar radio burst types, radio spikes are likely the shortest in duration, with narrow spectral widths.
They are emitted over wide frequency ranges, and suggested to be caused by plasma emission \citep[e.g.][]{1975A&A....39..107Z, 1977SvA....21..612C,2014SoPh..289.1701M} or electron cyclotron maser (ECM) emission \citep[e.g.][]{1982ApJ...259..844M, 2011ASSL..375.....C,2011ApJ...743..145C} depending on the coronal conditions at the height of emission. 
Considering the narrow bandwidths with the plasma emission hypothesis implies 
that the production of Langmuir waves and stimulation of radio emission occurs over short distances previously suggested to be the result of weak electron beams---i.e. small spatial sizes and beam densities \citep{1972A&A....17..267T,2014SoPh..289.1701M}. 
Previous studies have considered spike emission resulting from electron acceleration occurring in many small sites \citep{1982A&A...109..305B} and the fragmentation of flare energy release \citep{1985SoPh...96..357B}.
At GHz frequencies, millisecond spikes are suggested to be produced via electron-cyclotron maser (ECM) emission in numerous sources due to magnetic inhomogeneities \citep{2008ApJ...681.1688R} that could produce fragmentation secondary to that of primary energy release \citep{1998PhyU...41.1157F}. However, it is difficult to reconcile the ECM emission model with observations below 130 MHz in post-eruption loop systems \citep{2011ApJ...743..145C} as the plasma density would be too high to satisfy the condition $f_\mathrm{ce}\gtrsim f_\mathrm{pe}$.

Whilst spikes have been observed and studied over a number of decades, imaging observations are relatively sparse and previously limited to decimeter wavelengths, relating spikes to the site of energy release \citep{1995A&A...302..551K,1997A&A...317..569K,2001A&A...371..333P}, near loop tops \citep{2002A&A...383..678B}, and at the site of magnetic loop compression induced by a CME \citep{2006A&A...457..319K}.

The earliest one-dimensional imaging observations of spikes at $2.8$~GHz \citep{1991ApJ...369..255G} have found that different spike sources 
are produced from the same one-dimensional location to within $0.9$~arcsec using amplitude 
and phase data, with a $\sim25$~arcsec displacement from a continuum source. \cite{1995A&A...303..249A,1996ApJ...469..976A} report linear source sizes up to $46$~arcsec at $5.7$~GHz, deconvolved from the instrument beam. They find the spikes to coincide spatially with an underlying microwave burst within $12$~arcsec. From two dimensional imaging, \cite{1995A&A...302..551K} resolve only the minor axis of the emission with an observed size of $90$~arcsec at $333$~MHz. More recently, \cite{2021ApJ...922..134B} measure a FWHM size from a 2D Gaussian fit of $20$~arcsec at $1.1$~GHz, however the burst is not spatially resolved due to a beam size of $60$~arcsec, and can only be considered as a lower limit. Generally, the majority of past imaging observations provide limited information of the events, without any specific detail of individual spike source evolution in time, frequency, and space. Clearly, improved spike imaging from high sensitivity instruments is required. 

At decameter wavelengths, spike analysis was restricted to dynamic spectra \citep[e.g.][]{1994A&A...286..597B,2014SoPh..289.1701M,2016SoPh..291..211S} until recent observations presented time and frequency resolved imaging of individual spike sources \citep{Clarkson_2021}. 
Since radio waves are subject to significant refraction and scattering effects as they propagate through a turbulent corona, causing extended time profiles, substantially larger source sizes, and displaced positions in the plane-of-sky \citep[e.g.][]{2017NatCo...8.1515K}, the unprecedented resolution of the LOw Frequency ARray (LOFAR) allowed tracking of the spike source evolution at fixed frequencies over sub-second scales. The results imply that radio-wave scattering governs the time duration and peculiar motion. Further, the observed characteristics of decameter spikes are consistent with that of individual Type IIIb striae in the same event including the same sense of polarization, suggesting a common physical mechanism in a region of space where the magnetic field strength is not sufficient to satisfy the conditions for ECM emission. Moreover, the bursts were weakly polarized, contrary to the strong polarization expected from ECM emission \citep{1982ApJ...259..844M}. As such, the decameter spikes were considered to be produced via plasma emission.

These results are consistent with simulations of radio-wave scattering \citep{Kontar_2019} that show a degree of anisotropy $\alpha$ between $0.2-0.3$ within the density fluctuation spectrum is required in order to explain observational results \citep{Kontar_2019,2020ApJ...905...43C,2020ApJ...898...94K,2021A&A...656A..34M}, 
where $\alpha\equiv q_\parallel/q_\perp$ and $q_{\parallel,\perp}$ are the wavevector components of the electron density fluctuations. The simulations in \cite{2020ApJ...898...94K} show that sources observed away from the disk centre can present superluminal centroid velocities, 
as was observed for striae \citep{2020A&A...639A.115Z}, and spikes \citep{Clarkson_2021}. 
If the source motion is non-radial, then sources closer to the disk centre can also present speeds near $c$, as observed by drift-pair bursts \citep{2019A&A...631L...7K}.

In this work, we present analysis of over 1000 spikes using LOFAR, allowing a much needed statistical determination of the various spike characteristics between $30-45$~MHz from imaging observations and $30-70$~MHz from dynamic spectra, and compare with 250 individual striae of Type IIIb bursts. We analyse the centroid locations of both spikes and striae within a closed magnetic loop in conjunction with a CME, and compare the results to both scattering simulations, and observations over a wide frequency range within the literature.

\section{Overview of the Observations}

\begin{figure*}[htb!]
    %\figurenum{text}
    \epsscale{1.2}
    \plotone{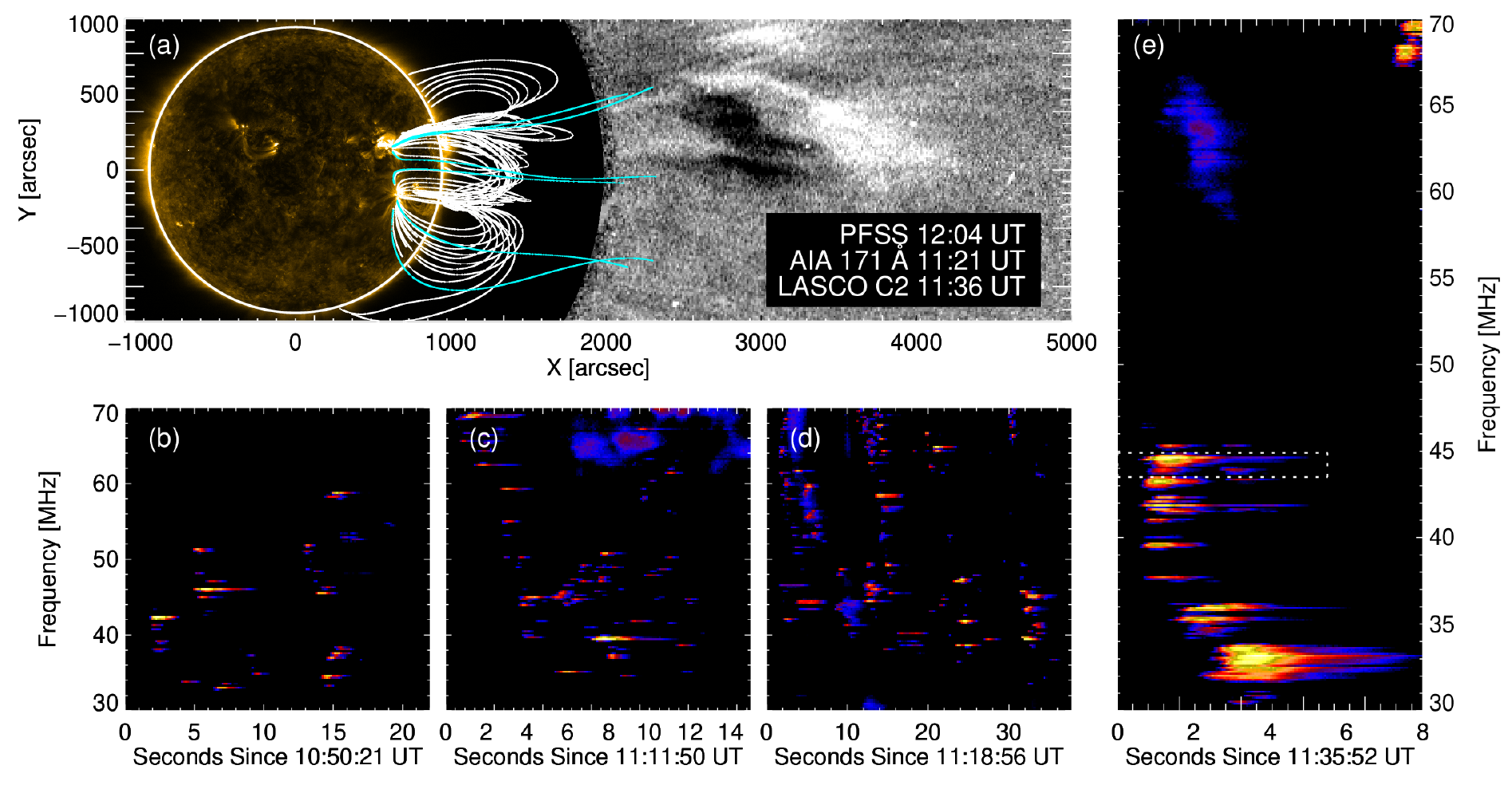}
    \caption{An overview of the event. \textbf{(a)} SDO/AIA $171$ \r{A} image at 11:20:57 UT, superimposed with a potential field source surface (PFSS) extrapolation at noon showing both open (blue) and closed (white) field lines within region surrounding the active region and northern sunspot, and a LASCO C2 image showing the streamer-puff (above) and narrow (below) CME fronts at 11:36:05 UT, as shown in \cite{2020ApJ...893..115C}. \textbf{(b-d)} Dynamic spectra showing samples of the spike emission. \textbf{(e)} Dynamic spectra of a Type IIIb J-burst. The dotted box highlights the region shown in Figure \ref{fig:typeIIIb_spikes_centroids_comp}. All dynamic spectra are background subtracted where the background is defined using a region at the start of each dynamic spectra containing no bursts at all frequencies.}
    \label{fig:event_overview}
\end{figure*}

Clusters of radio spikes and Type IIIb bursts were observed by the LOw Frequency ARay (LOFAR; \citealp{2013A&A...556A...2V}) on 2017-July-15 between 10:17 to 11:39 UT, associated with a coronal mass ejection. Figure \ref{fig:event_overview}(a) shows the LASCO C2 field of view at 11:36 UT where two narrow CME fronts, one of which is a streamer-puff CME \citep{2005ApJ...635L.189B}, are observed to erupt near 10:52 UT as the result of a jet \citep[see][for details]{2020ApJ...893..115C}. Also shown is an SDO/AIA 171 \r{A} image of the Sun at 11:21 UT, and a PFSS extrapolation showing open and closed magnetic geometry at noon. The active region AR12665 responsible for the flaring emission is located within the western solar hemisphere at a longitude of $\theta=52\arcdeg$ and latitude $\phi=-8\arcdeg$ as viewed from Earth. 1076 individually resolved spikes and 250 striae have been analysed with 421 spikes (207 striae) at frequencies between $30-45$~MHz. Above this frequency, significant side lobe emission can often be brighter than the main-lobe and were thus not used for imaging. Figure \ref{fig:event_overview}(b-d) shows example dynamic spectra of the spike emission. Six Type IIIb J-bursts within the event are chosen for analysis, one of which is shown in Figure \ref{fig:event_overview}(e) with a starting frequency of $45$~MHz for the fundamental, with a possible harmonic branch observed between $60-66$~MHz correlating to the lower frequency portion of the fundamental.

The observations were designed to have tied-array beams between $30-70$~MHz using the Low Band Antenna (LBA) stations in the outer configuration, with a maximum core baseline of $\sim 3.6$~km. 
The configuration provides 217 interferometrically synthesized beams that image the solar corona up to a radius of $\sim 3$~R$_\odot$, with a maximum temporal and spectral resolution of $10$~ms and $12.2$~kHz, respectively \citep[see][for details]{2017NatCo...8.1515K}. To minimise noise, the data were reduced temporally to $20$~ms resolution. During the observation, a single beam at the northern outer edge of the mosaic pattern across the solar disk recorded no data, and was discarded during analysis. The beam intensities are interpolated on a regular grid to produce the radio images. Observations of Tau A were used to calibrate the flux \citep[e.g.][]{2017NatCo...8.1515K} to solar flux units (sfu; 1 sfu = $10^{-22}$~W m$^{-2}$ Hz$^{-1}$).

Spike and striae characteristics are measured from both dynamic spectra and imaging observations as in \cite{Clarkson_2021}. Specifically, the radio source (see Figure \ref{fig:typeIIIb_spikes_centroids_comp}, for an example) that is convolved with the LOFAR point-spread function (PSF) is well approximated by a 2D elliptical Gaussian that dictates the centroid position $x_c, y_c$ of the observed sources. The peak time is defined at the central index of the flux profile within 15\% of the maximum, with the rise and decay position at the start and end of the FWHM duration. Ionospheric refraction can cause a noticeable shift in the observed radio emission dependent on the source elevation. We correct all centroid positions with respect to the zenith angle at $z=32.4\arcdeg$ for average ionospheric conditions as in \cite{2022ApJ...925..140G} (see equation 6).

%===================================================================

\section{Acceleration Region}

\subsection{Type IIIb Bi-Directional Exciter Motion}

\begin{figure*}[htb!]
    %\figurenum{text}
    \epsscale{0.9}
    \plotone{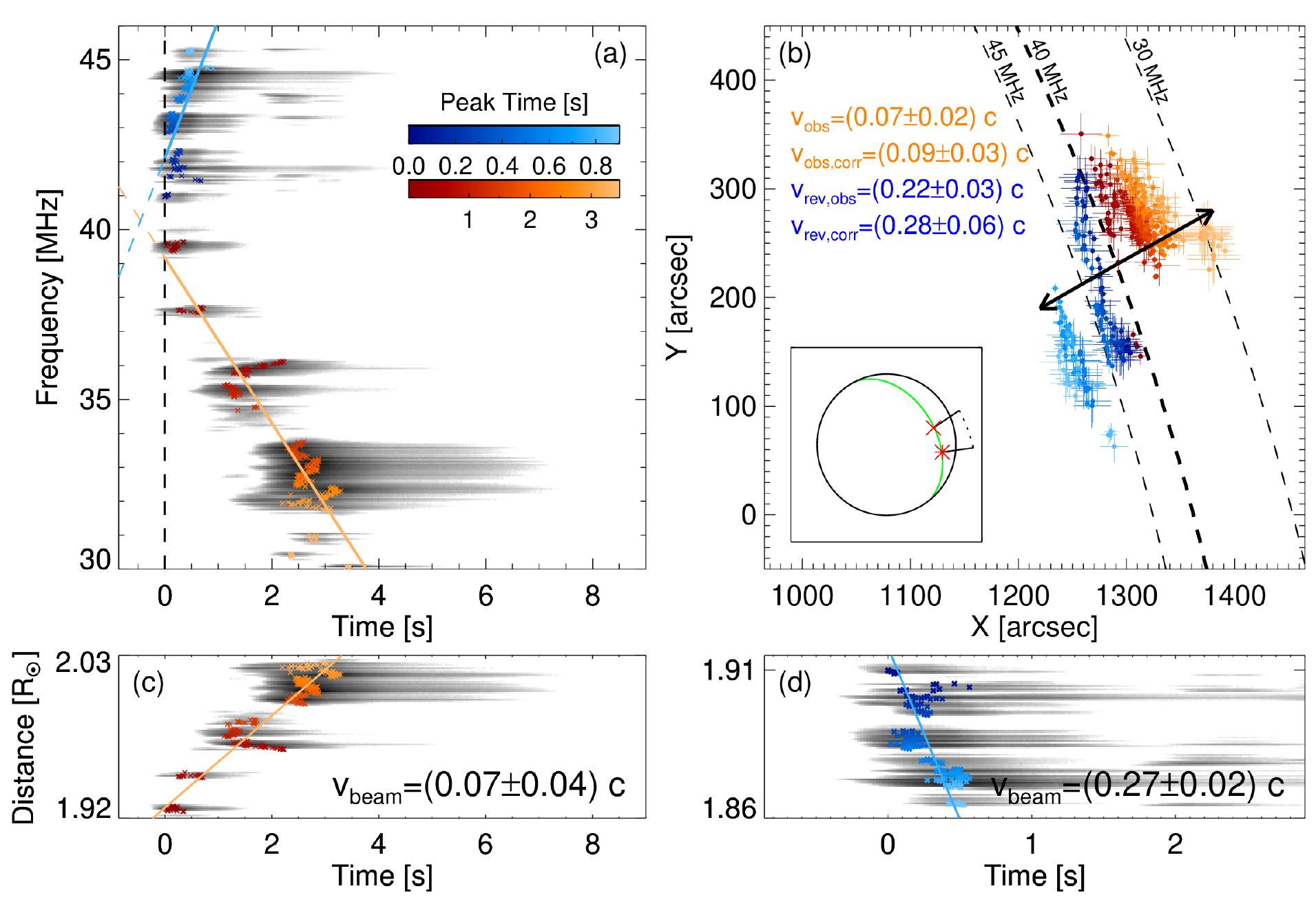}
    \caption{Centroid positions at the striae peaks of the Type IIIb J-burst, marked by the coloured points in the dynamic spectrum of panel (a). The orange and blue lines in panel (a) highlight the opposite sign bulk drift rates. In all panels, the orange points relate to the negative drifting burst segment, and the blue points relate to the reverse-drifting section. The colour gradients show the elapsed time from the earliest striae peak, represented by the vertical black dashed line in panel (a). Panel (b) shows the centroid locations of each striae with the arrows representing the observed trajectories and radial distances used to estimate the centroid velocities $v_\mathrm{obs}, v_\mathrm{rev}$ of each component across the plane-of-sky. A velocity estimate with a correction for the projection effect is given as $v_\mathrm{obs,corr}$ and $v_\mathrm{rev,corr}$ assuming the sources propagate along the sun-centre and active region plane with an uncertainty corresponding to a magnetic field spread of $30\arcdeg$. The black dashed curves show the exponential loop density model of equation \ref{eq:loop_ne_model} rotated as demonstrated in the inset of panel (b). Panels (c) and (d) show the distance-time spectra of each burst component with the distance obtained from the density model of equation \ref{eq:loop_ne_model}. The gradient of the linear fits to the striae peaks provides the beam velocities.}
    \label{fig:typeIIIb_centroids_time}
\end{figure*}

The Type IIIb bursts provide a useful diagnostic of the coronal loop that their sources partially trace. The example shown in Figures \ref{fig:event_overview}(e) and \ref{fig:typeIIIb_centroids_time}(a) has a negative bulk drift of $-6.1$~MHz s$^{-1}$ near $40$~MHz, comparative to previously reported Type IIIb drift rates \citep{2018SoPh..293..115S,2018ApJ...856...73C}, and reduces to $-1.1$~MHz s$^{-1}$ at $32$~MHz as the exciter approaches the loop apex, similar to other decameter J-bursts \citep{2017A&A...606A.141R}. Above these frequencies, the burst exhibits a reverse-drift at $4.9$~MHz s$^{-1}$ signifying electrons that are propagating towards the footpoints. Figure \ref{fig:typeIIIb_centroids_time} shows the peak centroid locations for the J-burst as a function of time, separated by the bulk drift direction. The bi-directional burst exciter motion \citep{1995ApJ...455..347A,2016SoPh..291.2407T} suggests that the acceleration region of the Type IIIb is at a radius corresponding to $\sim40$~MHz where the opposing centroid motions (bulk drifts) emanate, implying that in general, the site of acceleration for this event is high in the corona.

\subsection{Inferred Loop Density Model}

Figure \ref{fig:typeIIIb_centroids_time}(b) displays the frequency position according to an exponential loop density model \citep[e.g.][]{1999ApJ...515..842A} given by
\begin{equation}\label{eq:loop_ne_model}
    n_e(l)=A_n\exp{\left(-\frac{r(l)}{r_n}\right)},
\end{equation}
where $A_n=10^{11}$~cm$^{-3}$, $r_n=1.57\times10^{10}$~cm is density scale height similar to values found by \cite{2017A&A...606A.141R} for Type III U and J bursts, and $r(l)$ is the height along the loop from the solar surface. The angle formed between the latitude and longitude of each active region to the solar north is $\Phi\sim-27\arcdeg$. The density model is then rotated around the $x$ and $y$ axes as
\begin{equation}\label{eq:rot}
\begin{split}
    x^\prime &= x\cos{\theta}\\
    y^\prime &= y\cos{\Phi} + x\sin{\theta}\sin{\Phi},
\end{split}
\end{equation}
that produces the dashed lines representing the frequencies of the density model. The value of $r_n$ was then chosen to best represent the centroid locations at a given frequency. The inset of Figure \ref{fig:typeIIIb_centroids_time}(b) displays the applied rotation.

\subsection{Type IIIb Beam Velocities}\label{section:T3_beam_vel}

The centroid positions of the reverse-slope component of the Type IIIb J-burst are distributed across $80.6\pm11$~arcsec of the sky-plane over $0.9$~s. The distance uncertainty is derived from the average centroid uncertainties $\overline{\delta{x_c}},\overline{\delta{y_c}}$ as $\delta{r}=\left[(\partial{r}/\partial{x}\overline{\delta{x_c}})^2 + (\partial{r}/\partial{y}\overline{\delta{y_c}})^2\right]^{1/2}$. This correlates to a beam velocity of $v_\mathrm{rev,obs}=(0.22\pm0.03)c$. For the negative drift component, the observed radial spread is $103.0\pm9$~arcsec over $3$~s, giving $v_\mathrm{obs}=(0.07\pm0.02)c$. Due to the projection effect, these observed values are likely to be larger. The actual velocities can be estimated as $v_\mathrm{obs}/\sin{52\arcdeg}$ assuming that the sources propagate along the plane connecting the sun-center and active region. The uncertainty in this velocity is increased since we do not know the spread in angle that the magnetic field geometry makes with the active region. Here we assume a spread of $30\arcdeg$. The projection corrected beam velocities are then $v_\mathrm{corr}=(0.09\pm0.03)c$ and $v_\mathrm{rev,corr}=(0.28\pm0.06)c$. The beam velocities can also be estimated via the assumption of a density model (equation \ref{eq:loop_ne_model}). Panels (c-d) of Figure \ref{fig:typeIIIb_centroids_time} show the distance-time spectra of each burst component where the striae peaks are fit with a linear model. The gradient then provides the beam velocity through space as $(0.07\pm0.04)c$ and $(0.27\pm0.02)c$ for the  normal and reverse drifting components, respectively. Whilst each method to determine the beam velocity requires some assumption (projection effect or density model), the determined velocities agree within their errors.

\begin{figure*}[htb!]
    %\figurenum{text}
    \epsscale{1.2}
    \plotone{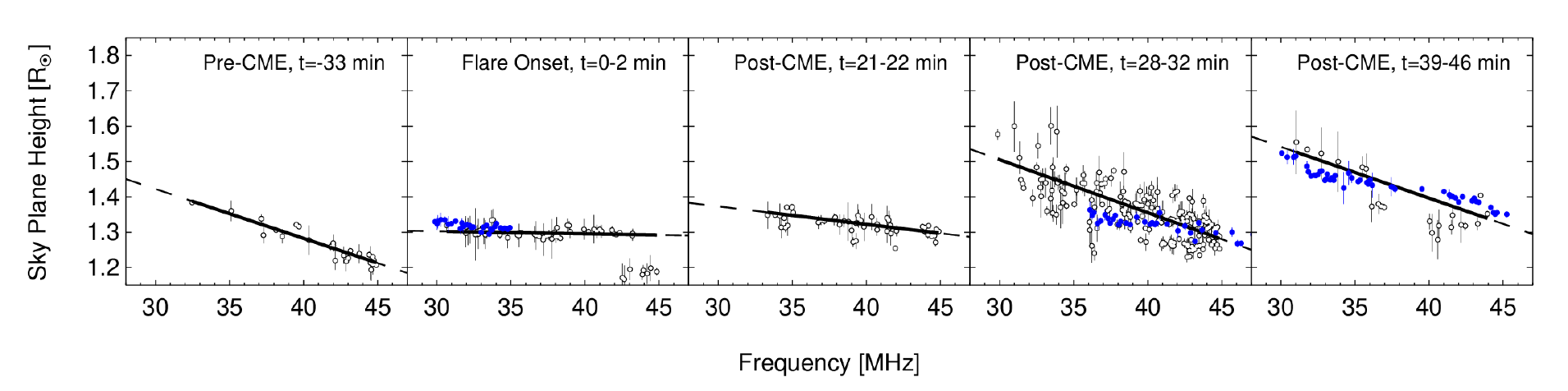}
    \caption{Observed centroid radial heights with frequency in the sky-plane for spikes (black, open circles) and striae (blue, closed circles). Each panel shows bursts that are close in time, with the second panel near the time of the flare, such that the prior panel shows pre-CME spikes. The spike data are fit with linear models.}
    \label{fig:spike_t3b_heights}
\end{figure*}

\begin{figure}[htb!]
    %\figurenum{text}
    \epsscale{1.1}
    \plotone{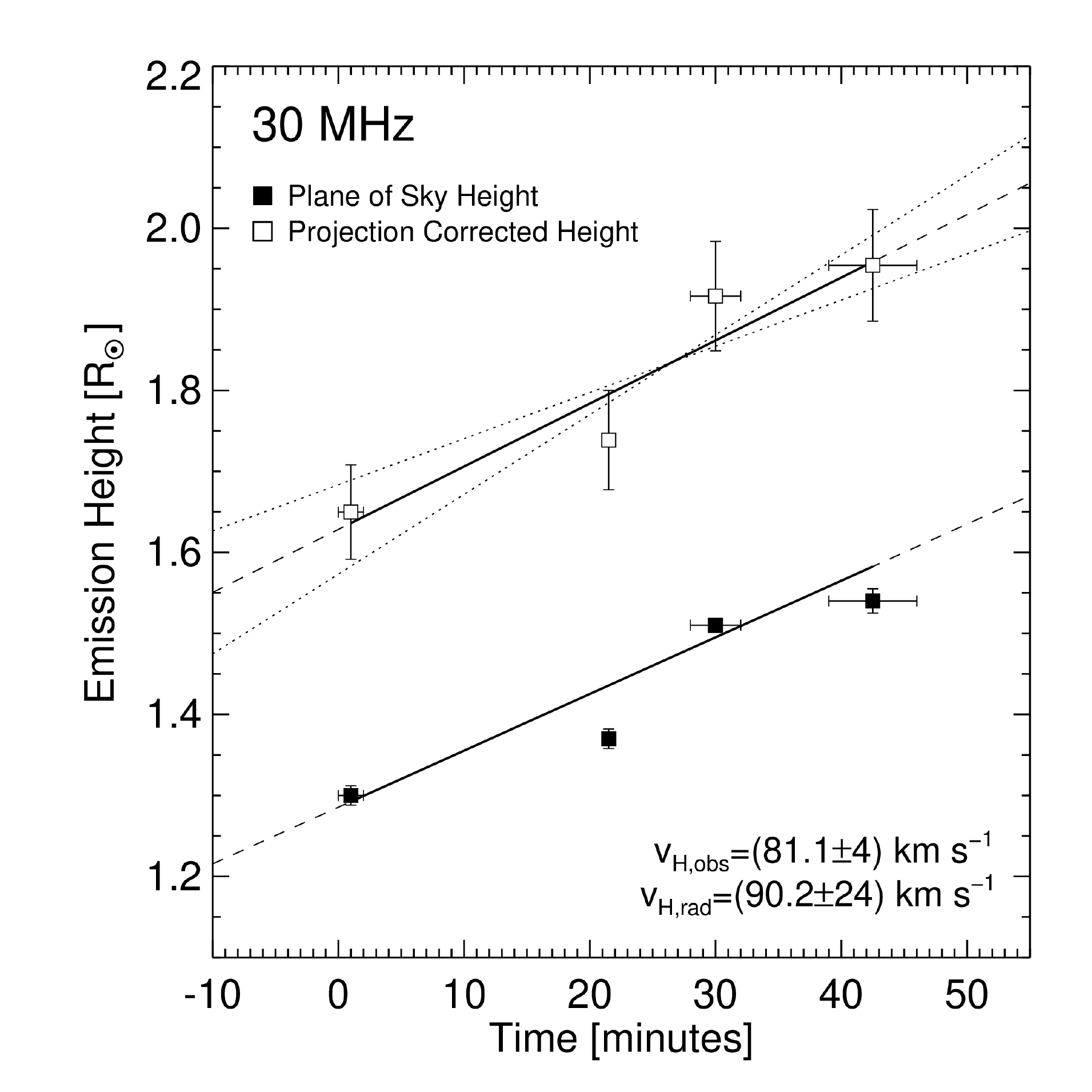}
    \caption{Observed sky-plane heights (solid squares) and estimated radial heights of the spike bursts over time at $30$~MHz extrapolated from the linear fits in Figure \ref{fig:spike_t3b_heights}. The horizontal error denotes the width of the time interval in each panel of Figure \ref{fig:spike_t3b_heights}. The vertical error of the plane-of-sky heights is derived from the linear fits in Figure \ref{fig:spike_t3b_heights}. The projection corrected heights assume a spread in angle of the loop magnetic field of $30\arcdeg$, contributing to the increased uncertainty. These data are fit with a linear model (solid lines) where the gradient describes the velocity at which the emission height at $30$~MHz increases over time. The dotted lines demonstrate the fit uncertainty for the projection corrected heights.}
    \label{fig:spike_30MHz_heights}
\end{figure}

\subsection{Acceleration Site Location}

Figure \ref{fig:spike_t3b_heights} shows the observed spike and striae centroid heights in the sky-plane as a function of frequency and time. For the bursts pre-CME, the heights vary linearly with frequency between $30-45$~MHz. After the onset of the flare and passage of the CME, the heights at a given frequency are reduced. In panel 2, the variation with frequency is minimal and then progressively increases over $\sim45$~minutes. The increase in the observed sky-plane height at $30$~MHz over time is shown in Figure \ref{fig:spike_30MHz_heights} from $1.30-1.54$~R$_\odot$. The variation is linear, and correlates to a velocity of $v_\mathrm{H,obs}=(81\pm5)$~km s$^{-1}$. The actual radial heights can be estimated by correcting for the projection effect as presented in section \ref{section:T3_beam_vel}, leading to a corrected velocity of $v_\mathrm{H,corr}=(90.2\pm24)$~km s$^{-1}$.

%===================================================================

\section{Observed Characteristics}

\subsection{Comparison of Spike \& Striae Centroid Positions}

\begin{figure}[htb!]
    %\figurenum{text}
    \epsscale{1.1}
    \plotone{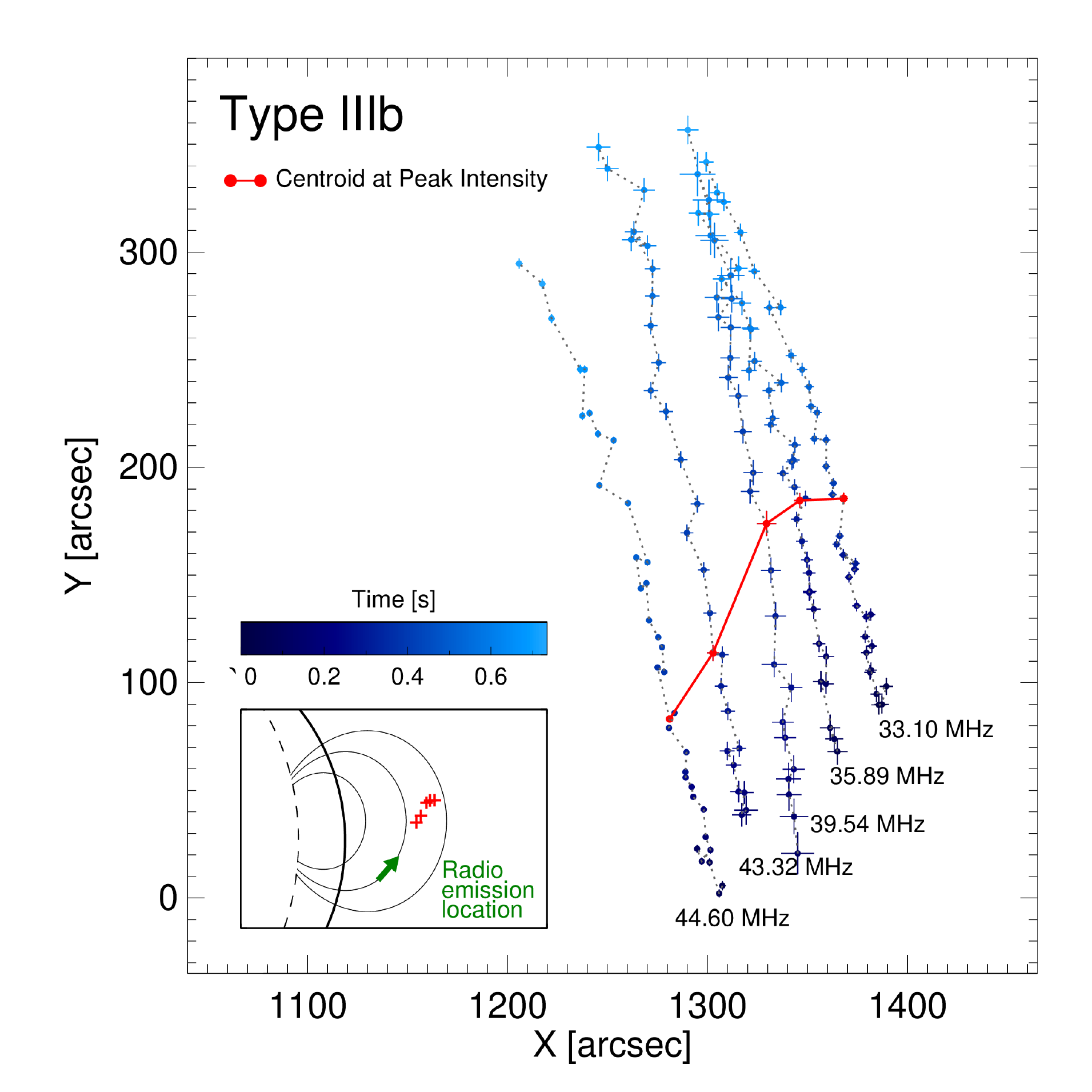}
    \caption{Type IIIb J-burst centroid locations of individual striae throughout the FWHM of the time profile at fixed frequencies. The red points and connected line represent the centroid at the peak intensity of each striae. The inset depicts coronal loop field lines with the likely location and path of the radio source (green arrow) corresponding to where the peak trajectory matches the field line direction. The red plus symbols mark the peak centroid locations.}
    \label{fig:typeIIIb_spikes_centroids}
\end{figure}

The centroid position of both spikes and striae differs greatly depending on the point in time that is used to generate the radio image. Figure \ref{fig:typeIIIb_spikes_centroids} shows centroid motion across the FWHM of the time profile of individual striae from the Type IIIb at various fixed frequencies. There are two visible components of motion---a frequency drift away from the Sun associated with the path of the exciter, and a displacement over time at fixed frequencies almost parallel to the solar limb towards the solar north, which is typically increased during the decay phase compared to the rise phase. We note that whilst the frequency drift and exciter trajectory does not correlate to the local field line direction of an inferred coronal loop, it implies that the actual emitter location is farther down the loop leg in a region where the field line trajectory matches the observed frequency drift direction, as indicated by the green arrow in the inset of Figure \ref{fig:typeIIIb_spikes_centroids}.

\begin{figure*}[!htb]
    \minipage{0.363\textwidth}
      \includegraphics[width=\linewidth]{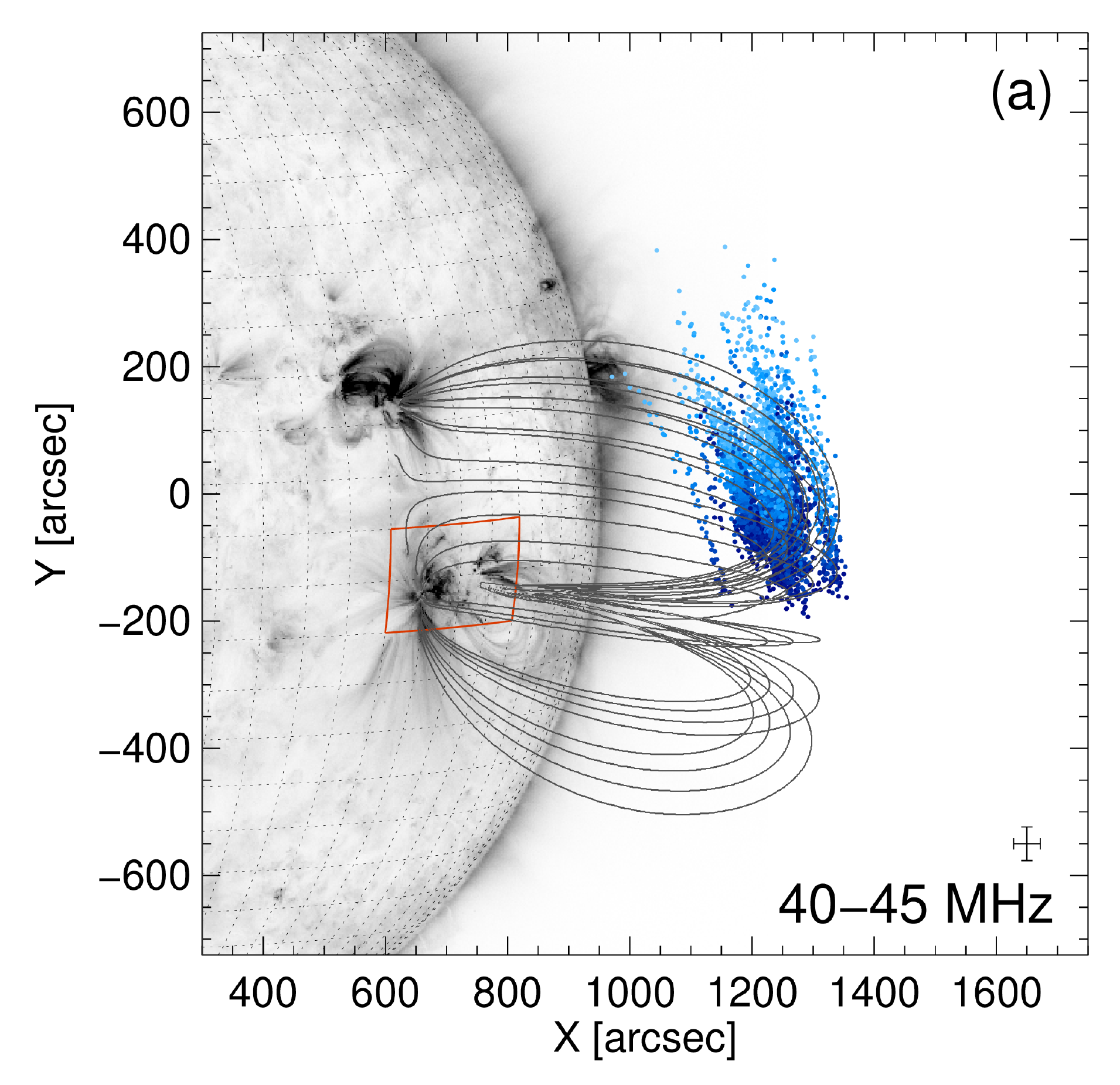}
    \endminipage\hfill
    \minipage{0.31\textwidth}
      \includegraphics[width=\linewidth]{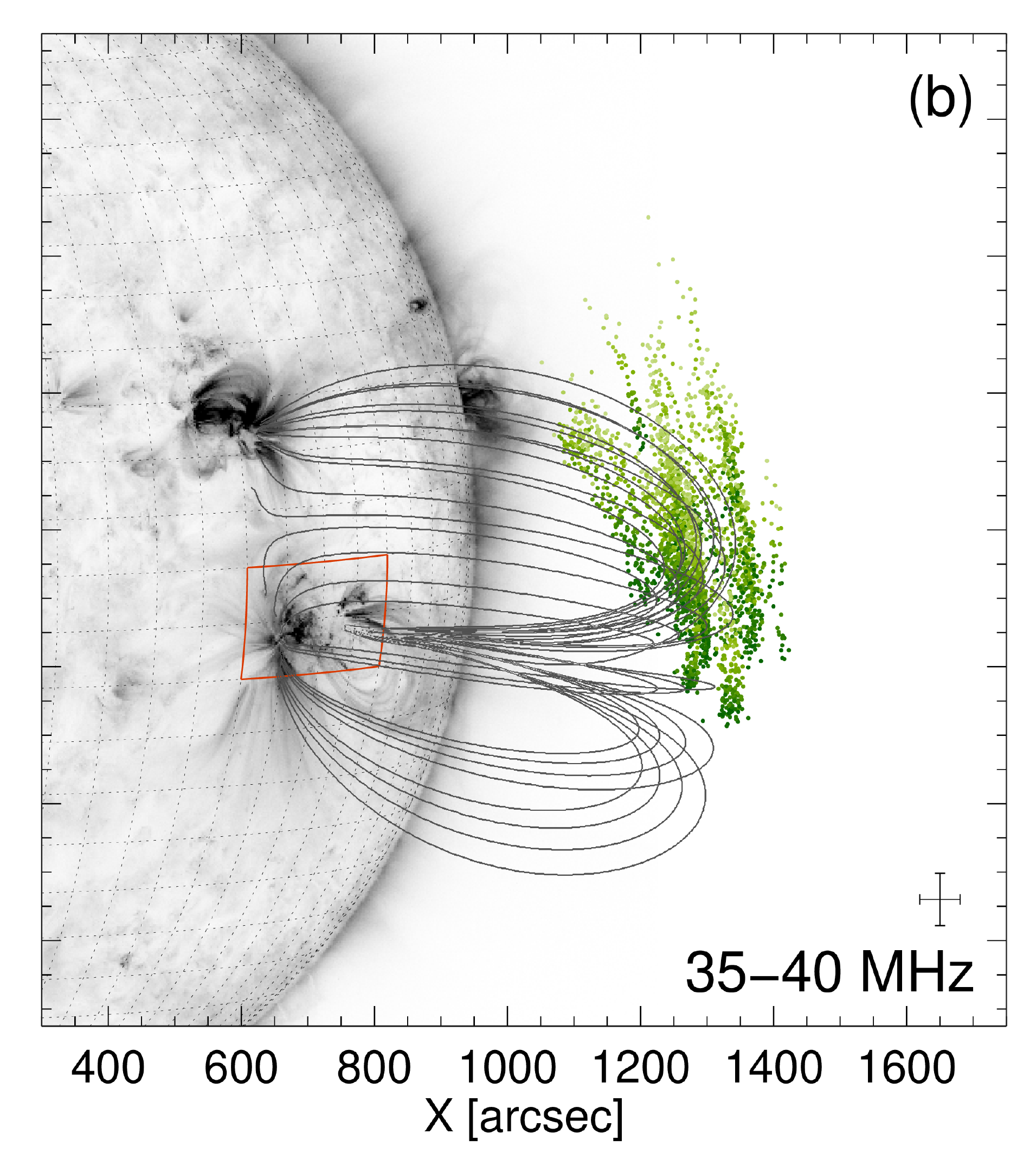}
    \endminipage\hfill
    \minipage{0.31\textwidth}%
      \includegraphics[width=\linewidth]{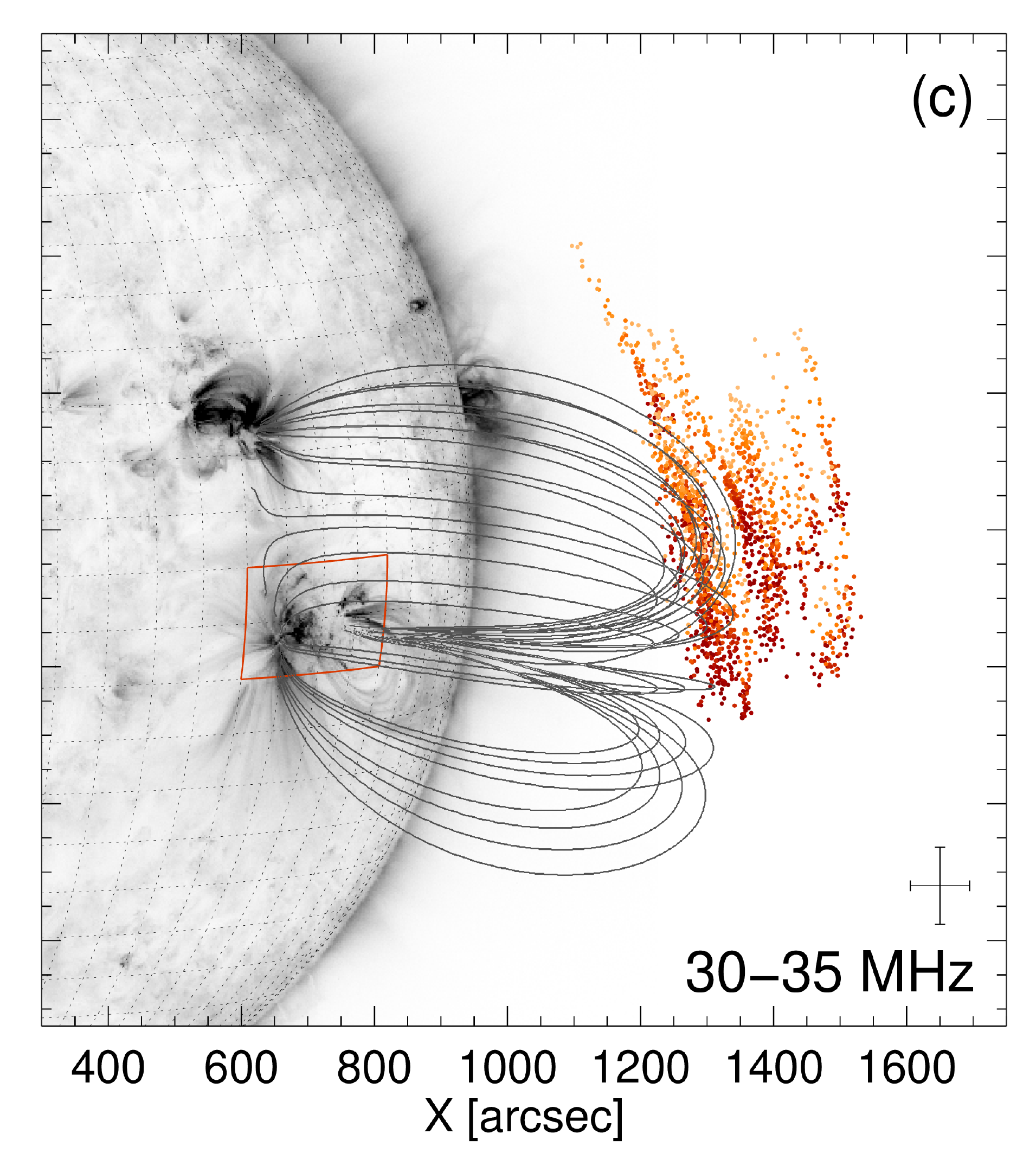}
    \endminipage
    \caption{Centroid positions of post-CME spikes across the FWHM intensity. Each individual spike is measured along its central frequency, with the collective motion grouped between \textbf{(a)} $40-45$~MHz, \textbf{(b)} $35-40$~MHz, and \textbf{(c)} $30-35$~MHz, and overlaid on an SDO/AIA image at 171 \AA. The colour gradients represent time increasing from dark to light. The average centroid error within each frequency band is indicated at the bottom right of each panel. The thin grey lines show the closed magnetic field lines from a PFSS extrapolation. The red box on the solar surface bounds the active region.}
    \label{fig:spike_motion}
\end{figure*}

The fixed frequency centroid motion evolution of spikes across the FWHM time duration are shown in Figure \ref{fig:spike_motion}. The individual and collective motion exhibits a displacement trajectory consistent with that shown by the striae, correlating in position with the closed magnetic field lines implied from the PFSS extrapolation. At a given frequency between $40-45$~MHz, the spikes and striae centroids appear to drift linearly, yet at lower frequencies there is a curvature in their motion.

\begin{figure*}[htb!]
    %\figurenum{text}
    \epsscale{1.15}
    \plotone{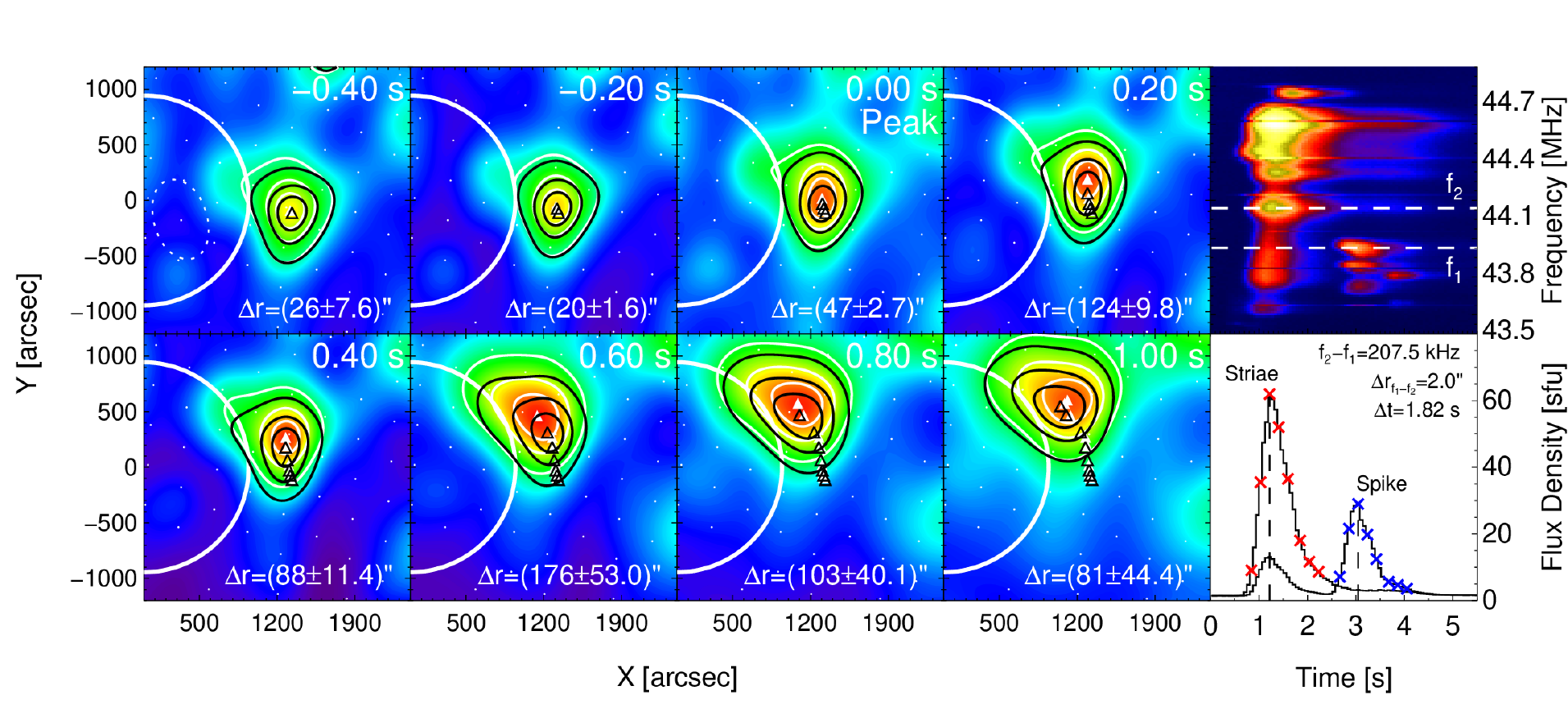}
    \caption{Comparison of spike and striae image contours with time. The left panels display contours of spike (white) and striae (black) emission from bursts at $44$~MHz that have a peak flux $1.82$~s apart, overlaid on the spike images with the background (average intensity of the faintest ten beams) subtracted. The contour levels are given at 90, 75 and 50\% of the maximum image flux. The peak of the each burst is set to $t=0.0$~s, with each panel showing the image at intervals of $0.2$~s before or after the peak. The triangle symbols track the motion of the spike (filled white) and striae (black) main lobes, with each symbol marking the average position of the image peak within $10\%$ of the maximum flux at each time. The separation distance of the peak locations in each panel is given by $\Delta{r}$ with the uncertainty given as $\delta{r}=\Delta{r}[(\Delta{x}_\mathrm{err}/\Delta{x})^2 + (\Delta{y}_\mathrm{err}/\Delta{y})^2]^{1/2}$ where $\Delta{x}_\mathrm{err},\Delta{y}_\mathrm{err}\propto \delta{I}/I_\mathrm{0}$, and the peak intensity $I_\mathrm{0}$ uncertainty is $\delta{I}\sim 1$~sfu. The white dots show the LOFAR beam locations, with the dotted oval in the first panel representing the beam size. The right two panels show the dynamic spectrum with the time profiles of the striae and spike sources along the white dashed lines. The red and blue crosses mark the time positions correlating to the images. $\Delta{r}_\mathrm{f_1-f_2}$ represents the distance in space between $\sim200$~kHz according to the density model of equation \ref{eq:loop_ne_model}.}
    \label{fig:typeIIIb_spikes_centroids_comp}
\end{figure*}

Figure \ref{fig:typeIIIb_spikes_centroids_comp} compares an individual spike and striae separated by $1.82$~s and $207$~kHz. The contours represent the intensity of the radio main-lobes, overlaid on the apparent images of the spike emission. The burst contours are aligned in time at their respective peaks, and each panel shows the contour locations in intervals of $0.2$~s before and after the peak. During the rise phase, the contour $90\%$ levels overlap almost entirely with the average peak intensity locations above the $90\%$ level separated by tens of arcsec. At the $50\%$ level, the spike contours are varied in shape compared to the striae towards the top left of the image. This is likely a secondary lobe of the PSF \citep[see Figure 6d in][]{2022ApJ...925..140G} and the lower intensity of the spike emission. During the decay phase, the contours are shifted vertically and rotated towards the limb by a similar angle, with the intensity peaks becoming separated. The shifted distance shown here is greater than that displayed in Figures \ref{fig:typeIIIb_spikes_centroids} and \ref{fig:spike_motion} as the latter only show the FWHM period.

\subsection{Spike Centroid Velocity, Source Area, \& Expansion}

\begin{figure}[htb!]
    %\figurenum{text}
    \epsscale{1.1}
    \plotone{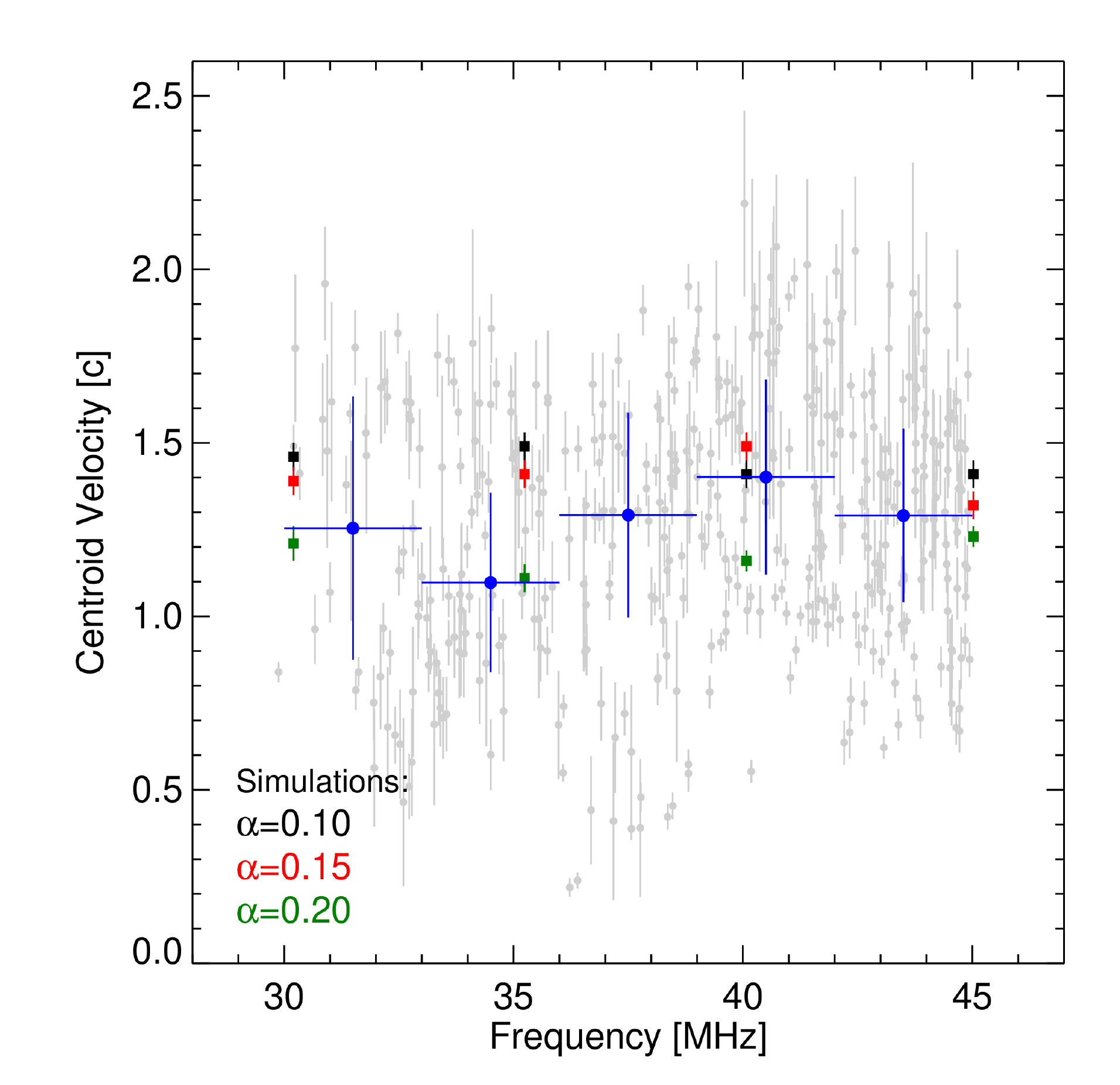}
    \caption{Spike plane-of-sky centroid velocities. The light gray points show the data with associated uncertainties. The blue points show the median values across $3$~MHz bins with the vertical error as the interquartile range representing the 25th and 75th percentiles. The squares show the centroid velocity calculated from scattering simulations of an initial point source injection located at $\theta=52^\circ$, anisotropy of the density fluctuation spectrum between $\alpha=0.1-0.2$, and a density fluctuation variance of $\epsilon=0.8$.}
    \label{fig:vel_scat_sim_angle}
\end{figure}

From the observed centroid motion in time at fixed frequencies, the spike plane-of-sky centroid velocity can be measured as $v=\sqrt{v_x^2 + v_y^2}$, where $v_x=\mathrm{d}x_c/\mathrm{d}t$ is obtained from linear fits to the $x$ centroid position over time, and similarly for $v_y$. The velocities are often superluminal with an average of $1.27c$ (Figure \ref{fig:vel_scat_sim_angle}). The measured centroid velocities from radio-wave scattering simulations \citep{Kontar_2019,2020ApJ...898...94K} are consistent with the median observed spike velocities, and both data sets show no frequency dependence between $30-45$~MHz.

The area of the radio source is measured at the FWHM level given by $A=(\pi/4)S_\mathrm{maj}S_\mathrm{min}$ where $S_\mathrm{maj}$ and $S_\mathrm{min}$ are the FWHM major and minor axis sizes of the fitted ellipse. The errors are estimated as in \cite{2017NatCo...8.1515K}. At $30$~MHz, the angular resolution of LOFAR is $\sim9$~arcmin, and the beam size at the time of observation is $115$~arcmin$^2$, reducing to $64$~arcmin$^2$ at $45$~MHz. The areal and linear expansion rates are given by fitting the change in area and ellipse axes widths over time during the decay phase with a linear model. Individual spike characteristics measured from imaging can be seen in the appendix (Figure \ref{fig:imaging_char}).

\subsection{Spike Temporal \& Spectral Profiles}

The lightcurve of each spike was analysed at the central frequency $f_c$. Variation between individual burst time profiles gives ambiguity in approximating all bursts with a single model, so we measure the rise $\tau_r$ and decay $\tau_d$ times directly from the data between the first point above the half-maximum level to the peak ($\tau_r$) and from the peak to the final value above the half-maximum level ($\tau_d$). The FWHM spike duration 
is then $\tau=\tau_r + \tau_d$ \citep{1990A&A...231..202G,1994A&A...286..597B,2018A&A...614A..69R}. The uncertainty on the rise and decay times is a combination of the background ($\sim1$~sfu at $30$~MHz, rising to $3-4$~sfu at $70$~MHz) and flux uncertainty that is typically $\sim 1$~sfu.

The drift rate of each spike across the FWHM duration was calculated by fitting the frequency-flux profile at each time index with a Gaussian, and fitting the position in time and frequency of each Gaussian peak with a linear model where the gradient provides the spike frequency drift rate \citep[see][for an example]{Clarkson_2021}. For this event, the spike drift rates are a few tens of kHz per second, with a weak tendency to increase with frequency.

\begin{figure}[htb!]
    %\figurenum{text}
    \epsscale{1.1}
    \plotone{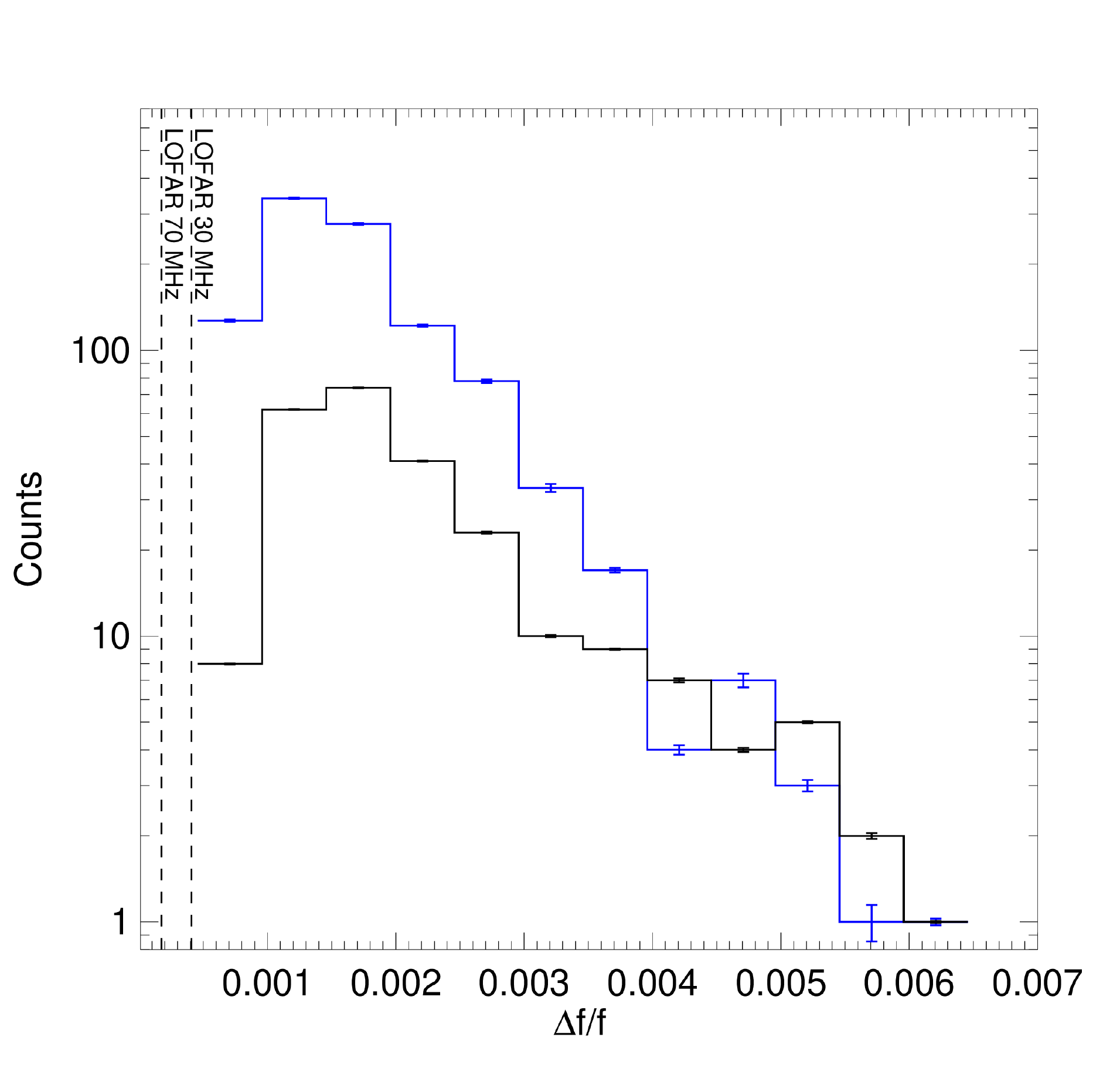}
    \caption{Histogram of the spike (blue) and striae (black) bandwidth ratio $\Delta{f}/f$ distributions with a bin size of $0.05\%$. The bin errors are given as $\delta{N}=N\left(\sum_i\delta{f_i^2}/\sum_i\Delta{f_i^2}\right)^{0.5}$, where $N$ is the bin count and $\delta{f}$ is the error on the measured bandwidth. The vertical dashed lines show the LOFAR frequency resolution ($12.2$~kHz) at $30$ and $70$~MHz.}
    \label{fig:spike_striae_bandwidth_dist}
\end{figure}

The spectral shapes of the spikes are symmetrical and well approximated by a Gaussian of form $I=I_0\exp{\left[-(f-f_c)^2/2\sigma^2\right]}+I_\mathrm{bg}$ where $I_\mathrm{bg}$ gives the background intensity, and $\sigma$ is the standard deviation. The FWHM spectral bandwidth is given by $\Delta{f}=\sigma\cdot2\sqrt{2\ln{2}}$. Figure \ref{fig:spike_striae_bandwidth_dist} shows the histogram of spike bandwidth ratios within this event. The distribution is asymmetrical, with a peak near $0.1\%$ followed by a tail that extends up to $0.6\%$. The peak is similar to that reported by \cite{2014SoPh..289.1701M} between $0.2.-0.3\%$ at $20-30$~MHz, yet lower than GHz observations by \cite{2008ApJ...681.1688R} at $0.7\%$, although this latter value is decreased to $0.5\%$ by \cite{2008ApJ...689..545N} using an updated spike deconvolution technique. Both aforementioned distributions at GHz frequencies extend up to $3\%$, a factor of five above that observed by LOFAR in this study, perhaps due to the employed algorithms ability to deconvolve overlapping spikes with larger bandwidths, or differing emission mechanisms. Figure \ref{fig:spike_striae_bandwidth_dist} also includes the bandwidth ratio for individual striae that share a similar peak and spread to the spikes in this event. The slopes of the two distributions are comparable at the smallest bandwidth ratios, whilst the trends differ above $\Delta{f}/f\approx0.0035$, potentially due to overlapping striae broadening the measured bandwidth at lower frequencies.

Individual spike characteristics measured from dynamic spectra can be seen in the appendix (Figure \ref{fig:ds_char}). We note that some characteristic averages show an oscillation with frequency that correlates with an oscillation in the observed flux from Tau A (Figure \ref{fig:ds_char}(a)). Therefore, we regard this as an instrumental artefact and not the result of a physical solar process.

\section{Comparison of Decameter and Decimeter Spike Observations}

\subsection{Image Sizes}

\begin{figure}[htb!]
    %\figurenum{text}
    \epsscale{1.1}
    \plotone{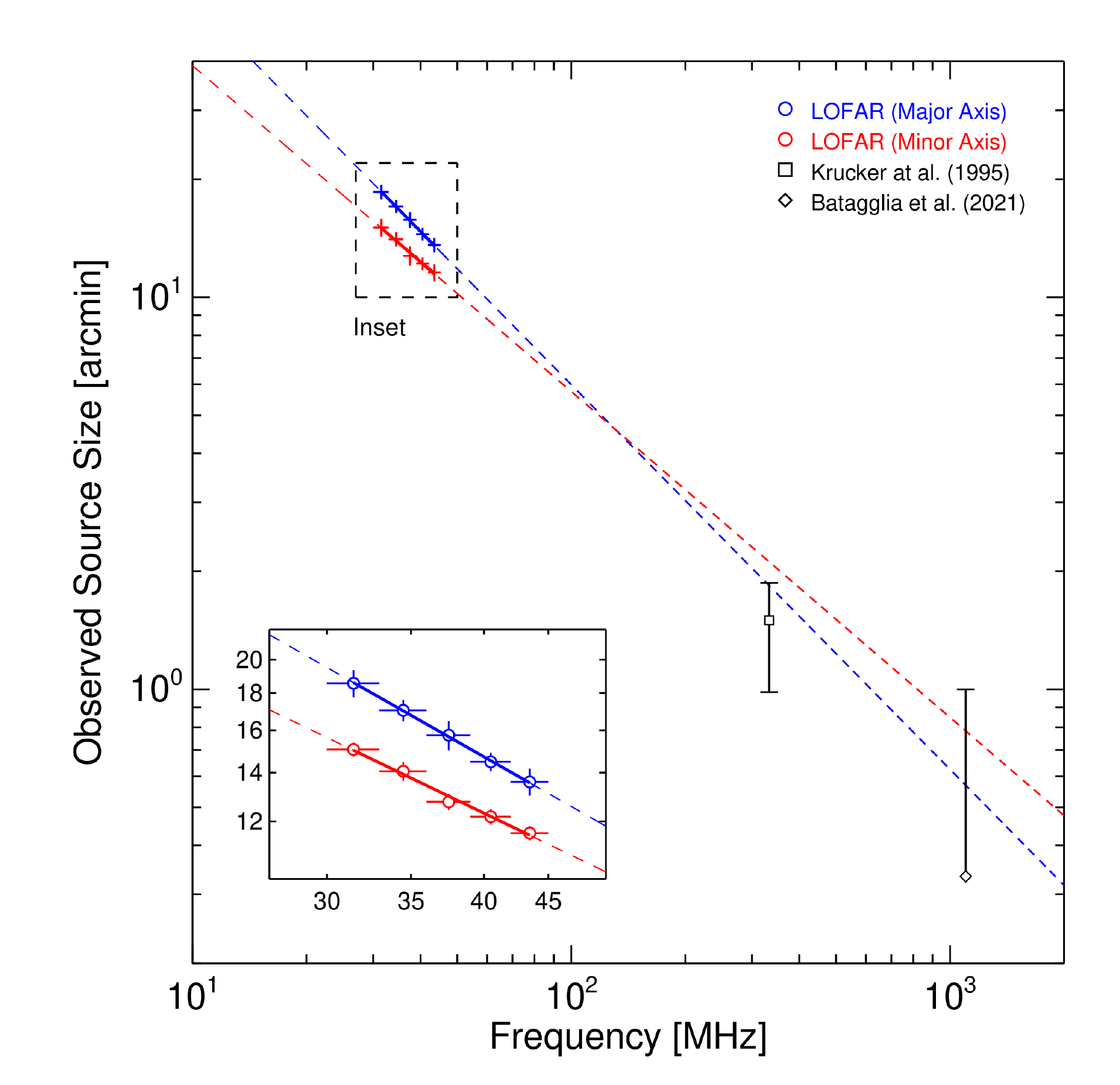}
    \caption{Observed source sizes of spikes. The blue and red points show the median major and minor axes FWHM sizes observed by LOFAR from Figure \ref{fig:imaging_char}(b,c). \cite{1995A&A...302..551K} observes the half-max contour sizes of $110$~arcsec (major) and $90$~arcsec (minor), yet the instrument beam is $112$~arcsec (major) and $59$~arcsec (minor), such that the source is resolved only along the minor axis---here we plot the minor axis size only, with the error showing the upper and lower sizes of the instrument beam. \cite{2021ApJ...922..134B} note that the bursts are not spatially resolved, such that the FWHM of the fitted 2D Gaussian is a lower limit. The error presented here shows the instrument beam size.}
    \label{fig:spike_size}
\end{figure}

The limited image measurements from previous studies \citep{1995A&A...302..551K,2021ApJ...922..134B} are compared with the median linear spike sizes observed by LOFAR in Figure \ref{fig:spike_size}. The trend is consistent between each frequency range, although the decimeter measurements are not resolved along the major axes. The LOFAR obtained major axis sizes range from $19.5$~arcmin at $30$~MHz to $13$~arcmin at $45$~MHz, decreasing as $f^{-0.98\pm0.2}$. The minor axis sizes are $\sim0.8$ times the major axis, decreasing as $f^{-0.83\pm0.1}$.

\subsection{Decay Times}\label{section:DecayTimes}

\begin{figure}[htb!]
    %\figurenum{text}
    \epsscale{1.1}
    \plotone{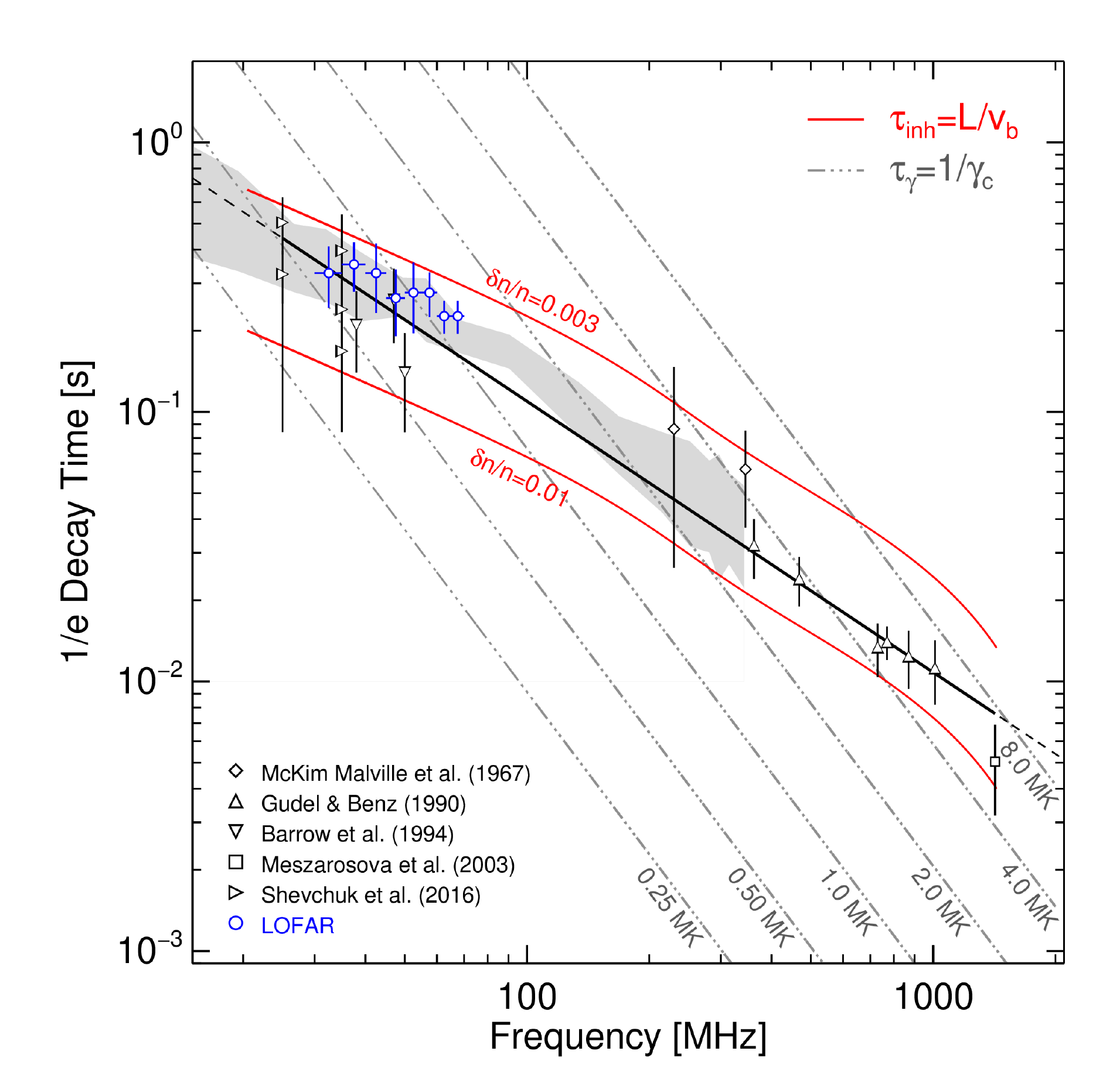}
    \caption{Average spike $1/e$ decay times against frequency. LOFAR data show the median from Figure \ref{fig:ds_char}(c) adjusted from the FWHM by a factor of $\sqrt{\ln{2}}$. The solid black line represents a power-law fit to the data given by equation \ref{eq:decay_freq}. The grey dash-dotted lines represent the plasma collision time for various coronal temperatures. The grey region shows the scattering decay time for $\alpha=0.2$ between $\epsilon=0.5$ (lower bound) and $2.0$ (upper bound). The red curves show the inhomogeneity time as defined by equation \ref{eq:inh_time} for fixed values of $\delta n/n$ as shown. For each curve, the inhomogeneity scale varies from $2$~km at $1.03$~R$_\odot$ to $100$~km at $2$~R$_\odot$.}
    \label{fig:spike_decay}
\end{figure}

Figure \ref{fig:spike_decay} combines spike average $1/e$ decay time measurements from several authors (\citealp{1967ApJ...147..711M, 1990A&A...231..202G, 1994A&A...286..597B, 2003A&A...407.1115M, 2016SoPh..291..211S}) between $25$~MHz and $1.42$~GHz. The linear fit in log-space indicates a $1/f$ dependence as
\begin{equation}\label{eq:decay_freq}
    \tau_d(f) = (11.22\pm1.9)f^{(-1.01\pm0.03)}~\mathrm{s},\quad 25\le f \le 1420~\mathrm{MHz}.
\end{equation}
We include the collisional damping time given as $\tau_\gamma=1/\gamma_c$ for various coronal temperatures $T_e$ where
\begin{equation}\label{eq:gamma_c}
    \gamma_c=\pi n e^4 z_p^2 \ln{\Lambda}/(m_e^2 v_{Te}^2),
\end{equation}
as defined in \cite{2011A&A...529A..66R} but with an additional term $z_p=1.18$ giving the average atomic number in the photosphere \citep{2011A&A...536A..93J}, and with the Coulomb logarithm $\ln{\Lambda}\approx23$ \citep{2011SSRv..159..107H}. The density is derived from the frequency as $n=(f\mathrm{[MHz]}/8.93\times10^{-3})^2$, and the parameters $e, m_e, v_\mathrm{Te}$ are the electron charge and mass, and thermal velocity, respectively.

Also shown is the inhomogeneity time \citep[e.g.][]{2001A&A...375..629K} as
\begin{equation}\label{eq:inh_time}
    \tau=\frac{|L|}{v_b}=\frac{2n_e(r)}{v_b}\left(\frac{\mathrm{d}n_e(r)}{\mathrm{d}r}\right)^{-1}=\frac{2n_e(r)}{v_b}\frac{\lambda}{\delta n_e}
\end{equation}
Here we use the density profile $n(r)$ presented in \cite{Kontar_2019} with an additional term $n_c(r)=n_0\exp{(-(r-1)/h_0)} + n_1\exp{(-(r-1)/h_1)}$ that accounts for a sharp increase in density towards the chromosphere, where $n_0=1.17\times10^{17}$~cm$^{-3}$ gives the density at the solar surface with $h_0=144$~km providing the density scale height \citep{2008A&A...489L..57K}, $n_1=10^{11}$~cm$^{-3}$, and $h_1=0.02$~km \citep[e.g.][]{2012ApJ...760..142B}. The final density is then $n_e(r)=n(r)+n_c(r)$. We set the beam velocity to $v_b=10^{10}$~cm s$^{-1}$, and choose the scale $\lambda$ to vary linearly from $2$~km at $1.03$~R$_\odot$ ($f_\mathrm{pe}=1.4$~GHz) to $100$~km at $2$~R$_\odot$ ($f_\mathrm{pe}=20$~MHz), with fixed density fluctuation amplitudes of $\delta n_e/n_e=0.003$ and $0.01$. This gives $|L|\sim (0.04-2)\times10^4$~km when $\delta{n_e}/n_e=0.01$ and $|L|\sim (0.1-6)\times10^4$~km when $\delta{n_e}/n_e=0.003$, where the upper values of $|L|$ are similar to that considered by \cite{2001A&A...375..629K} for electron beam dynamics with a fluctuating background density.

\subsection{Bandwidth}

\begin{figure*}[htb!]
    %\figurenum{text}
    \epsscale{1}
    \plotone{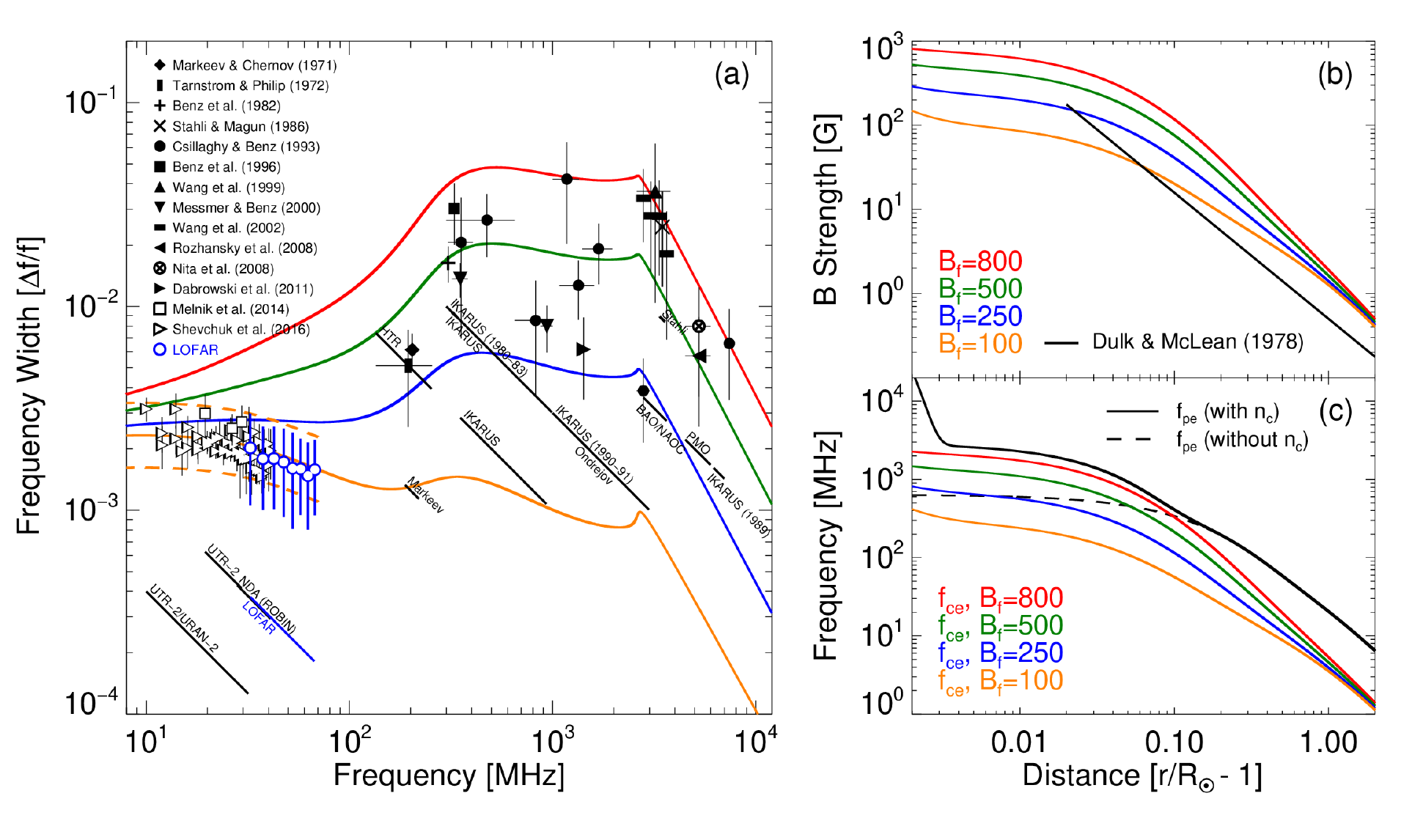}
    \caption{(a) Average spike bandwidth ratio $\Delta{f}/f$ (where $\Delta{f}$ is given at the FWHM level) against frequency combining observations as indicated in the legend. LOFAR data show the median from Figure \ref{fig:ds_char}(d). The diagonal lines represent the instrument resolutions. The coloured curves denote the bandwidth derived from the Langmuir wave dispersion relation in a magnetized plasma (equation \ref{eq:lwave_disp_rel}) with $\psi=23\arcdeg$. Each colour uses a the magnetic field model of \cite{2001SoPh..203...71G} with the constant $B_f$ varied. The orange dashed curves vary $\psi$ from $19\arcdeg$ (lower) to $28\arcdeg$ (upper). (b) The magnetic field models with distance. Also shown is the magnetic field model from \cite{1978SoPh...57..279D} as $B(r)=0.5(r/\mathrm{R}_\odot-1)^{-1.5}$. (c) The plasma and cyclotron frequencies from the density and magnetic field models.}
    \label{fig:fwidth_MHz_GHz}
\end{figure*}

Combined average spike bandwidth observations as a function of frequency are shown in Figure \ref{fig:fwidth_MHz_GHz} \citep{1971SvA....14..835M, 1972A&A....16...21T, 1982A&A...109..305B, 1985SoPh...96..357B, 1986SoPh..104..117S, 1992A&AS...93..539B, 1993A&A...274..487C, 1996A&A...309..291B, 1999SoPh..189..331W, 2000A&A...354..287M, 2008SoPh..253..133W, 2008ApJ...689..545N, 2008ApJ...681.1688R, 2002SoPh..209..185W, 2011SoPh..273..377D, 2014SoPh..289.1701M, 2016SoPh..291..211S, 2019ApJ...885...90T} from $10$~MHz to $8$~GHz. Despite the possibility of over-estimated average bandwidths at the higher frequencies due to limited spectral resolutions, the increase of $\Delta{f}/f$ above $200$~MHz is expected via consideration of the Langmuir wave dispersion relation in a weakly magnetized plasma \citep{1985srph.book..177M,2012wop..book.....P,2014SoPh..289.1701M,2016SoPh..291..211S} as
\begin{equation}\label{eq:lwave_disp_rel}
    \omega = \omega_\mathrm{pe} + \frac{3k^2v_\mathrm{Te}^2}{2\omega_\mathrm{pe}} + \frac{\omega_\mathrm{ce}^2}{2\omega_\mathrm{pe}}\sin^2{\psi},
\end{equation}
where $\psi$ is the angle between the plasma wave direction and the magnetic field, and $\omega_\mathrm{ce}=eB/m_e c$ is the electron cyclotron frequency. Equation \ref{eq:lwave_disp_rel} is valid under the condition that $\omega_\mathrm{ce}\ll\omega_\mathrm{pe}$ for electron beams where the spatial size is $<10^8$~cm \citep{2014SoPh..289.1701M}. Figure \ref{fig:fwidth_MHz_GHz} shows the expected bandwidth $\Delta\omega=\omega{(\psi)} - \omega(0)$ for varied magnetic field strengths based on the model by \cite{2001SoPh..203...71G} (see equation 2) with terms $B_f$ varied between $100-800$~G shown in panel (b) and the density model given by $n_e(r)$, shown in panel (c). A value of $B_f=500$~G gives a magnetic field strength of $\sim300$~G at a height of $0.02$~R$_\odot$ from the solar surface, as estimated by \cite{2017PhRvL.118o5101K}. We choose $\psi=23\arcdeg$ to provide the best match to the low frequency spike population trend when $B_f=100$~G, similar to the angle of $20\arcdeg$ by \cite{1975SvA....18..475Z} for Type III bursts.

%===================================================================

\section{Discussion}

We investigated 1076 solar radio spikes to statistically retrieve their characteristics in the decameter range, and compared their centroid locations with individual Type IIIb striae within a coronal loop structure. The apparent radio map contours of each burst type that are close in time (less than a few seconds) and frequency overlap with little variation in position and shape, with any clear separation occurring during the decay phase. Both burst types experience comparable and significant displacement along the loop direction at fixed frequencies with substantial broadening over time, predominantly along the major axis.

\subsection{Beam Velocities \& Emission Heights}

From the Type IIIb bulk frequency drift, the electron beam velocity can be estimated. The reverse-slope component corresponds to a beam velocity of $0.3c$, whilst the outward propagating beams has a velocity of $0.1c$, which could indicate an asymmetry in the energy injected in each direction.
Pre-CME, the spikes show an increase of height with decreasing frequency. After the CME ejection at approximately 10:52 UT (indicated by the Type III burst onset \citep{Clarkson_2021} and jet \citep{2020ApJ...893..115C}), the observed spike and Type IIIb striae heights weakly vary with frequency that could be caused by the magnetic field being distorted such that the density gradient and loop trajectory is along the line of sight. The loop may then restore over tens of minutes towards the prior configuration with a velocity of $\sim90$~km s$^{-1}$. Interestingly, the bursts near $45$~MHz seem to retain an approximately constant height, whilst those towards $30$~MHz experience the bulk of the shift, which implies the bulk of the rotation and/or expansion of the magnetic field occurs at the lower frequencies towards the loop top.

\subsection{Anisotropic Scattering}

The spike and striae centroids present two components of motion: an exciter-driven frequency dependent radial motion, and a shift perpendicular to the radial direction over time at fixed frequencies. The latter is assumed to be independent of the exciter motion since variations in the ambient density or magnetic field will be stable over the time periods of $0.1-0.5$~s. We attribute the fixed frequency displacement to radio-wave scattering in an anisotropic turbulent medium, that has previously been shown to produce such motion along the direction of the magnetic field \citep{2017NatCo...8.1515K,2020ApJ...898...94K}. The centroid locations can then provide insight into the magnetic field structure: above $40$~MHz, the displacement is linear, becoming arced towards the limb at lower frequencies such that the field structure may have greater curvature farther from the Sun.

We show that the median fixed frequency centroid velocities present no frequency dependence between $30-45$~MHz, and are replicated with radio-wave scattering simulations using strong levels of anisotropy ($\alpha=0.1-0.2$). Whilst varying the anisotropy factor causes a small change in the centroid velocity, the major influence is the magnetic field angle to the observer---for a field aligned perpendicularly to the observer's line of sight, the observed centroid velocity will be maximal. The spread in observed velocities at a given frequency for spikes at different times suggests that the source locations are distributed across a range of angles within the loop, and that the anisotropy and turbulence levels may fluctuate over time.

\subsection{Event Interpretation}

\begin{figure*}[htb!]
    %\figurenum{text}
    \epsscale{0.65}
    \plotone{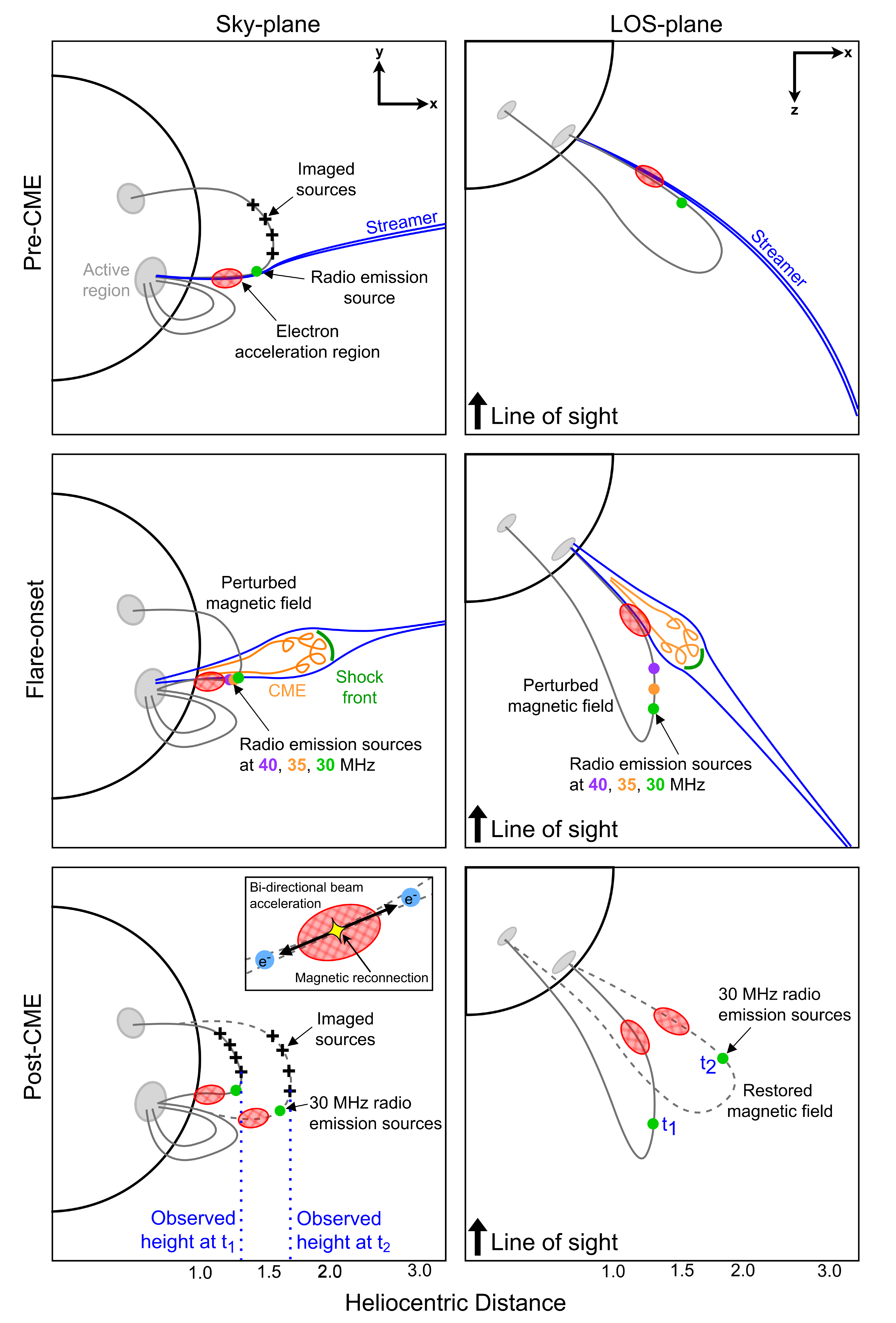}
    \caption{A cartoon of the event before and after the CME eruption. The two columns show the $x,y$ sky-plane and $z,x$ line-of-sight (LOS) planes, respectively, with each column showing the configuration pre-CME, during the flaring period, and post-CME. Pre-CME, a smaller number of spikes are observed with source emission locations (coloured circles) and electron acceleration (red hatched regions) likely occurring along the lower loop leg via magnetic reconnection interaction between the loops (grey lines) and streamer (blue lines). At the onset of the flare and eruption of the streamer-puff CME (orange line) caused by a jet at the lower active region, the streamer is inflated due to the CME-driven shock (green line) that also perturbs the magnetic loop geometry towards the observer. The rotation of the loop means that the frequency dependence of the observed spike sources is masked. Throughout the remaining observation window, repeated magnetic reconnection occurs, accelerating electron beams along the loop direction leading to an increased number of spike sources. In the post-CME phase, the magnetic field restores towards it's original configuration, such that the observed emission at the same frequency appears at larger sky-plane heights over time, moving across the sky-plane at a velocity of $90$~km s$^{-1}$ between $t_1$ and $t_2$. The imaged sources are observed to drift along the direction of the magnetic field due to anisotropic radio-wave scattering such that their centroids (black crosses) drift upwards in the sky-plane over time at fixed frequencies. The inset in the lower left panel shows counter-propagating electron beams induced by magnetic reconnection that can then produce the bi-directional Type IIIb. The streamer and shock front are not shown in the lower panels, and the smaller loops connected to the lower active region are excluded from the LOS-plane, for clarity.}
    \label{fig:cartoon}
\end{figure*}

Our interpretation of the event is shown in Figure \ref{fig:cartoon}. Beam acceleration most likely occurs along the ascending loop leg towards the lower flank of the CME, at a region closer to the Sun where the magnetic field geometry has a radial trajectory, inferred from the frequency drift in Figure \ref{fig:typeIIIb_spikes_centroids}. Interestingly, the Type II observed during this event is suggested to have occurred towards the opposite CME flank \citep{2020ApJ...893..115C}. The passage of the CME perturbs the magnetic geometry, rotating the loops towards the observer, causing any frequency dependence of emission to be masked in the sky-plane. Post-CME, repeated magnetic reconnection and subsequent electron beam acceleration produces a greater number of spikes than in the pre-CME case. As the field restores towards its original configuration, a sky-plane drift of the imaged emission source locations is observed at a speed of $90$~km s$^{-1}$. The distance from the acceleration region and location where the source begins emitting will depend on the beam density, velocity, and turbulence that determine the spatial location of Langmuir wave growth. The emission quickly undergoes radio-wave scattering, with the strongest scattering power perpendicular to the field lines at a given location, such that the likely direction of photon propagation is parallel to the field where the scattering power is weakest. This direction changes over time and distance due to the field geometry. The observed centroids are then displaced along the loop direction at fixed frequencies over time and expanded in area, with the true source location never observed.

We observed spikes spread across many frequencies, implying that the acceleration heights may vary, or the electron beams have differing initial densities and spatial sizes which can increase the onset time and burst starting frequencies \citep{2011A&A...529A..66R, 2018ApJ...867..158R}. The former agrees with the interpretation of spikes arising from many small sites of magnetic reconnection \citep{1982A&A...109..305B,1985SoPh...96..357B}, which may have been triggered by the CME. The frequency at which a spike is observed could then be an interplay between the acceleration location, the initial beam properties, and the coronal turbulence that will promote Langmuir wave growth at a specific region of space.

\subsection{Radio-wave Scattering Dominance}

The scatter-dominated characteristics of spikes (source size, decay time) present power-law trends between decameter and decimeter observations. At $30$~MHz, estimation of the intrinsic source size suggests $l<2r_n\Delta{f}/f\approx1$~arcsec where $\Delta{f}/f\sim2\times10^{-3}$. Following the same approach at $333$~MHz, \cite{1995A&A...302..551K} note a similar intrinsic size of $2$~arcsec using a density scale height of $10^{10}$~cm.
The spike duration and plasma collision time similarity has led to previous suggestions of this damping mechanism to be the dominant factor controlling their short time profiles \citep{1967ApJ...147..711M,1972A&A....17..267T,1986SoPh..104...99B,2014SoPh..289.1701M}. Using the collision time to infer the coronal temperature gives a range between $0.5-8.0$~MK from $25$~MHz to $1.4$~GHz. However, due to significant broadening, the inferred temperatures would be reduced. At $30$~MHz, typical decay times from simulations are $\sim0.2$~s for $\alpha=0.2$ \citep{Kontar_2019}, reducing the median observed decay time in this study to $0.1$~s and the coronal temperature to $<0.25$~MK, much lower than expected above an active region.
Figure \ref{fig:spike_decay} presents a $1/f$ trend for spike decay times, with a similar power-law index to that for Type III bursts \citep{Kontar_2019}. In addition, we show the inhomogeneity time $\tau_\mathrm{inh}$ which is the characteristic time for Langmuir waves to drift in velocity space \citep{2001A&A...375..629K}, and can match the observed $1/f$ trend for given input parameters as described in section \ref{section:DecayTimes}. We see that varying the value of $\delta n_e/n_e$ can explain the spread in the observed data, where larger fluctuation amplitudes causes the Langmuir waves to drift in velocity space faster, resulting in shorter decay times---and indeed, $\delta n_e/n_e$ would be expected to vary from event to event. However, the trend required to match the observations with frequency is found from the inhomogeneity length scale $\lambda$ linearly increasing with $r$---we estimate this to range from $2-100$~km between $1.03-2$ R$_\odot$. The inhomogeneity time is also similar to the scattering induced decay time from simulations (grey region in Figure \ref{fig:spike_decay}) which is dictated by the density fluctuations. Since the observed time profile is a combination of both the intrinsic duration and broadening due to scattering as $\tau=(\tau_\mathrm{source}^2 + \tau_\mathrm{scat}^2)^{1/2}$, if $\tau_\mathrm{source}\ll\tau_\mathrm{scat}$ as the observations suggest, then radio-wave scattering is the dominant contribution, and governs the observed decay time.

\subsection{Magnetic Field Strength}

The spread in $\Delta{f}/f$ can be predicted via the Langmuir wave dispersion relation (Figure \ref{fig:fwidth_MHz_GHz}), which is significantly modified near $400$~MHz through variation of the magnetic field strength which could vary substantially between events. We note that we do not describe the theoretical spectral shape of spikes produced via plasma emission with this relation, which remains an open question outside the scope of this study. For a fixed angle between the plasma wave and magnetic field of $\psi=23\arcdeg$ matching the average trend of the low frequency spikes (where a range of $\psi$ from $19-28\arcdeg$ encompasses the low frequency spike bandwidth ratios), the data suggest that between $0.4-3.5$~GHz the magnetic field strength between events could vary from $30-800$~G. In the decameter range, spikes observed between $20-70$~MHz ($1.45-2.0$~R$_\odot$ for the density model considered), the magnetic field strength varies between $1.2-3.5$~G, larger than the model above active regions by \cite{1978SoPh...57..279D} of $0.5-1.65$~G, however, this model is derived from a compilation of different techniques where the data have a spread within a factor of three. It suggests that decameter spike events are associated with active regions that have stronger magnetic fields than average by a factor of $\sim2$.

\section{Conclusion}

Solar radio spikes and Type IIIb bursts are observed to be associated with a trans-equatorial closed loop system, and could be associated with repeated magnetic reconnection in numerous small sites triggered by a streamer-puff CME. The typical observed bandwidth ratios suggest that the size of the emitting region is less than 1 arcsec, which evolves in position over several tens of minutes owing to a perturbed magnetic geometry caused by the CME and shock propagation. Spikes and striae that are close in time and frequency have $90\%$ intensity contours that almost completely overlap, suggesting the sources emit radio-waves from the same region of space. The emitting location will be determined by an interplay of the electron beam acceleration site, the beam characteristics, and the turbulent conditions. Fixed frequency imaging of both burst types reveals strongly directive, superluminal centroid motion along the guiding magnetic field parallel to the solar limb, consistent with radio-wave scattering in an anisotropic medium. The observed spread in centroid velocity could be due to varying anisotropy, turbulence level, and emission angle within the loop. The strong scattering environment means that the emitting source locations do not correspond to the locations of the observed sources---from the Type IIIb frequency drift, the beam trajectory is likely associated with a region of the loop closer to the Sun along the ascending leg where the field trajectory is closer to the radial direction. Consequently, the region of acceleration and emission could be near the CME flank. The frequency dependence of scattering dominated properties from decameter to decimeter wavelengths present a consistent $1/f$ trend, similar, but not identical to that for type III bursts \citep{2019ApJ...884..122K}, suggesting that radio-wave scattering is significant in both domains, and governs the observed decay and sizes. Assuming plasma emission, the observed spike bandwidth ratios can be replicated via the Langmuir wave dispersion relation for conditions where $f_\mathrm{ce}\ll f_\mathrm{pe}$, with an order of magnitude increase above $400$~MHz due to strongly varying magnetic field strength between events at this scale. However, ECM emission can not be discarded as the mechanism for spikes at GHz frequencies. In the decameter range, spike observations suggest that the magnetic field strength is stronger than average above active regions by a factor of $\sim2$.

\begin{acknowledgments}
    DLC, EPK, and NV are thankful to Dstl for the funding through the UK-France PhD Scheme (contract DSTLX-1000106007). EPK, XC are supported by STFC consolidated grant ST/T000422/1. NC thanks CNES for its financial support. We gratefully acknowledge the UK-France collaboration grant provided by the British Council Hubert Curien Alliance Programme that contributed to the completion of this work. The authors acknowledge the support by the international team grant (\href{http://www.issibern.ch/teams/lofar/}{http://www.issibern.ch/teams/lofar/}) from ISSI Bern, Switzerland. This paper is based (in part) on data obtained from facilities of the International LOFAR Telescope (ILT) under project code LC8\_027. LOFAR \citep{2013A&A...556A...2V} is the Low-Frequency Array designed and constructed by ASTRON. It has observing, data processing, and data storage facilities in several countries, that are owned by various parties (each with their own funding sources), and that are collectively operated by the ILT Foundation under a joint scientific policy. The ILT resources have benefited from the following recent major funding sources: CNRS-INSU, Observatoire de Paris and Universit\'{e} d'Orl\'{e}ans, France; BMBF, MIWF-NRW, MPG, Germany; Science Foundation Ireland (SFI), Department of Business, Enterprise and Innovation (DBEI), Ireland; NWO, The Netherlands; The Science and Technology Facilities Council, UK; Ministry of Science and Higher Education, Poland.
\end{acknowledgments}
\bibliographystyle{aasjournal}
\bibliography{refs}

\appendix
\section{Spike Characteristics with Frequency}
\begin{figure*}[!htb]
    \minipage{0.32\textwidth}%
      \includegraphics[width=\linewidth]{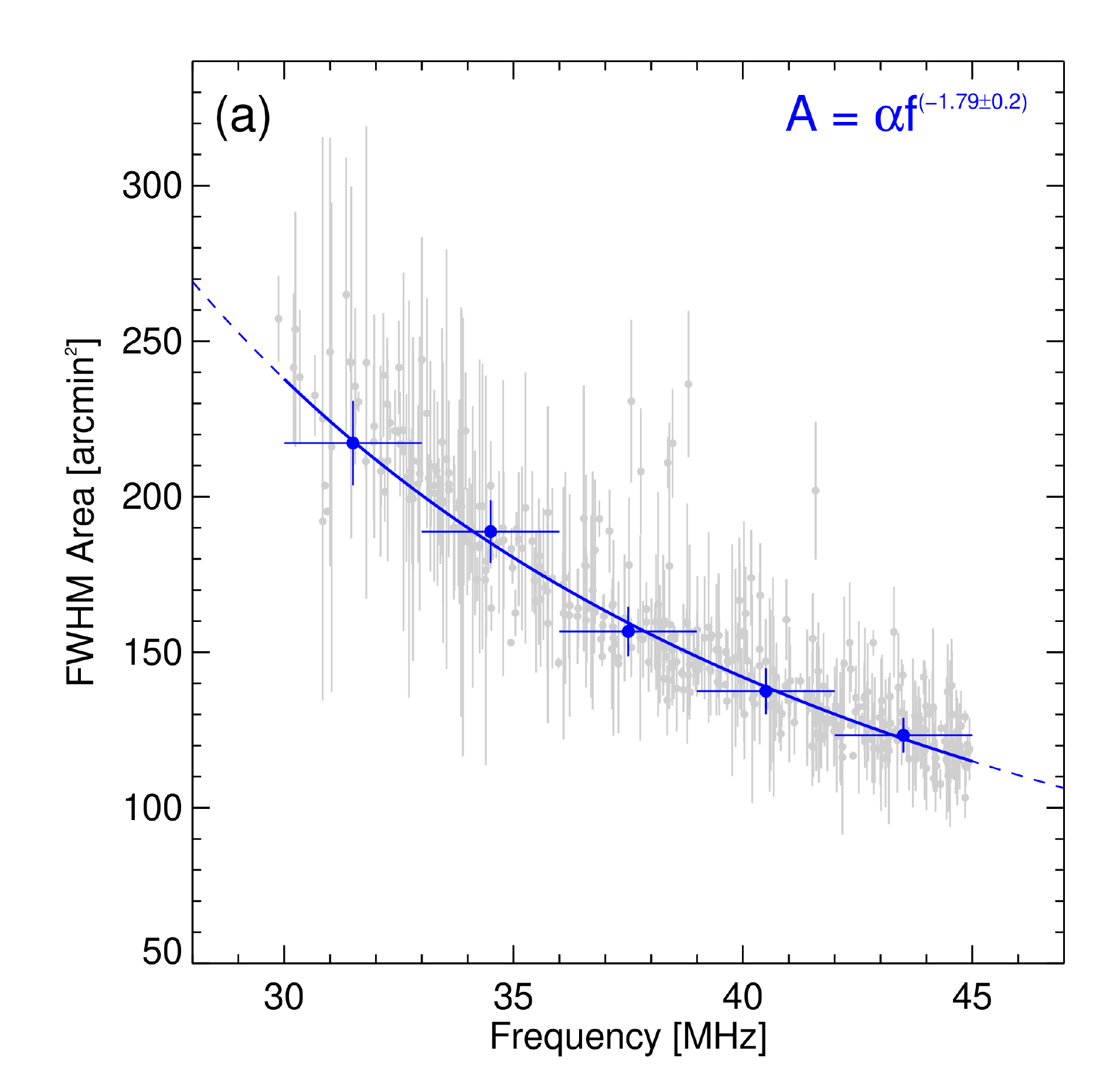}
    \endminipage\hfill
    \minipage{0.32\textwidth}%
      \includegraphics[width=\linewidth]{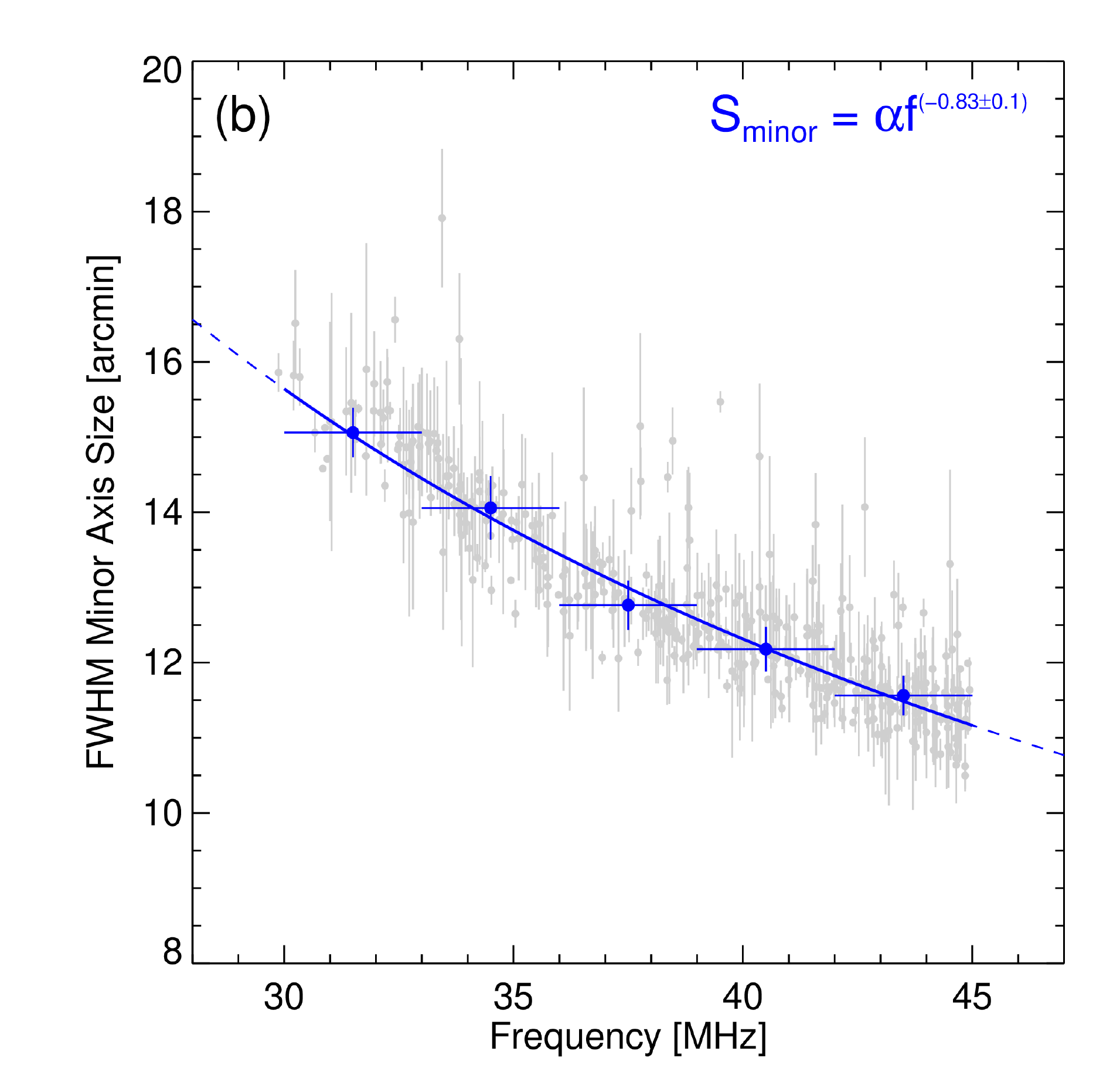}
    \endminipage\hfill
    \minipage{0.32\textwidth}%
      \includegraphics[width=\linewidth]{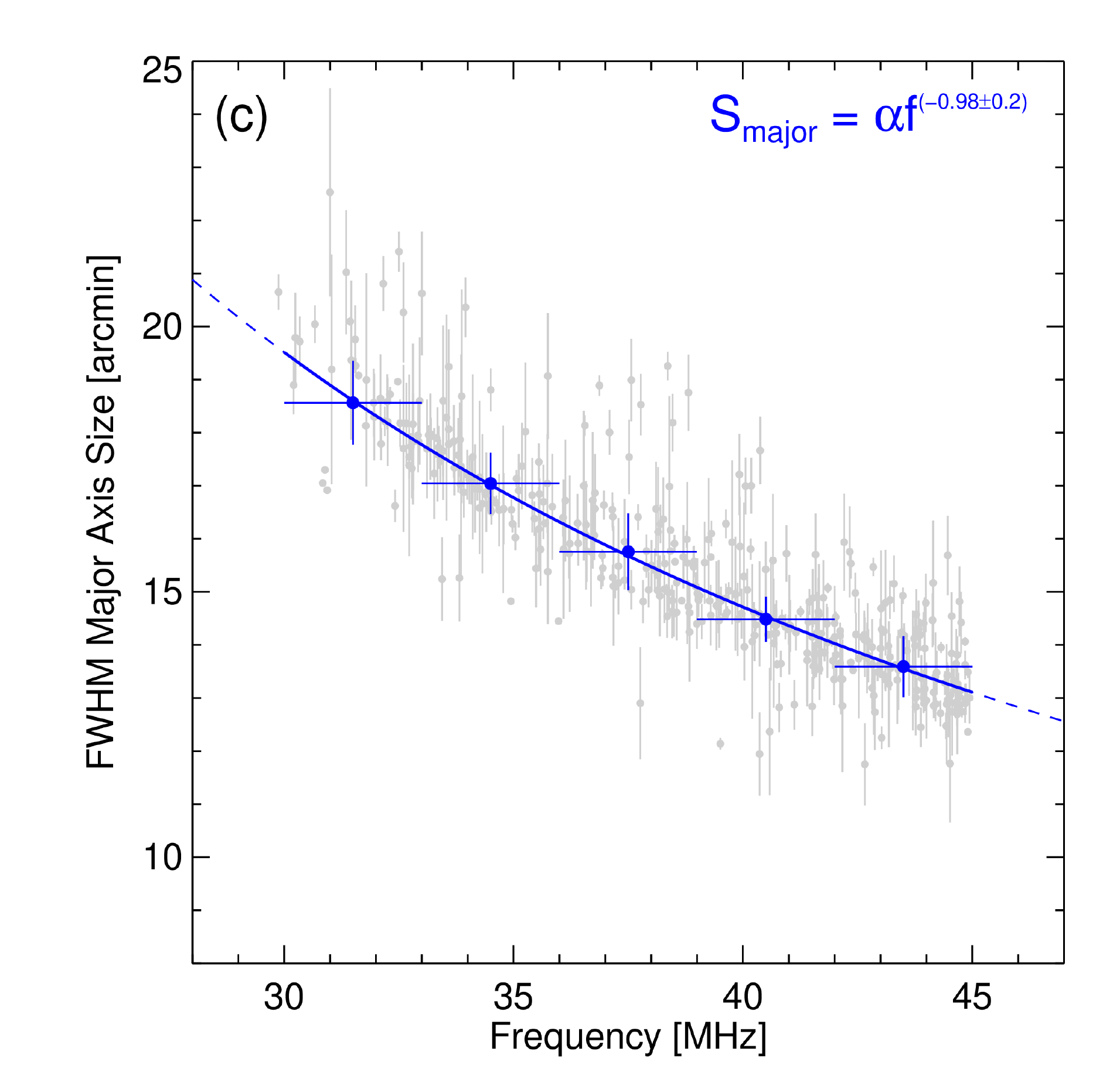}
    \endminipage\hfill
    \minipage{0.32\textwidth}%
      \includegraphics[width=\linewidth]{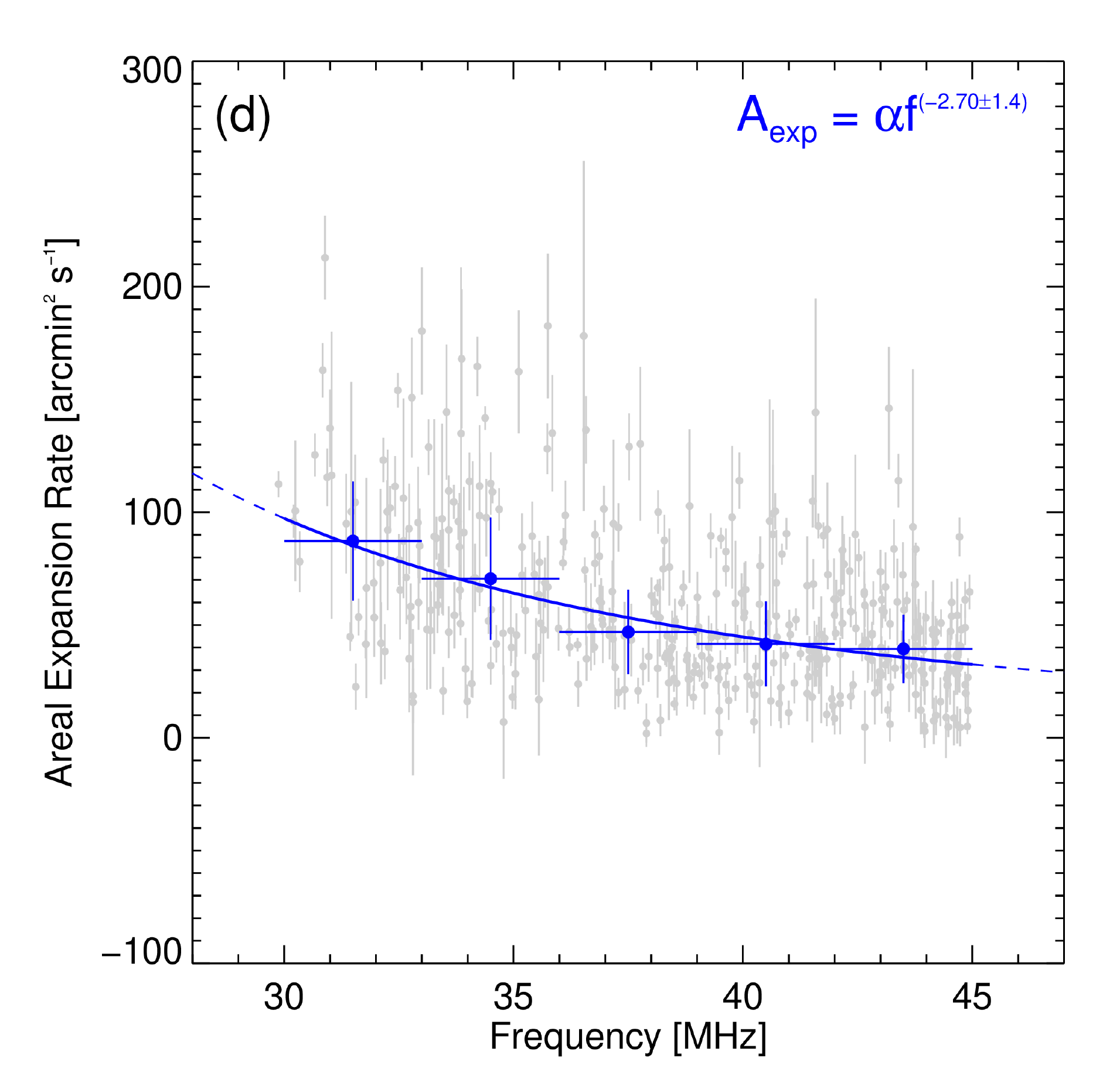}
    \endminipage\hfill
    \minipage{0.32\textwidth}%
      \includegraphics[width=\linewidth]{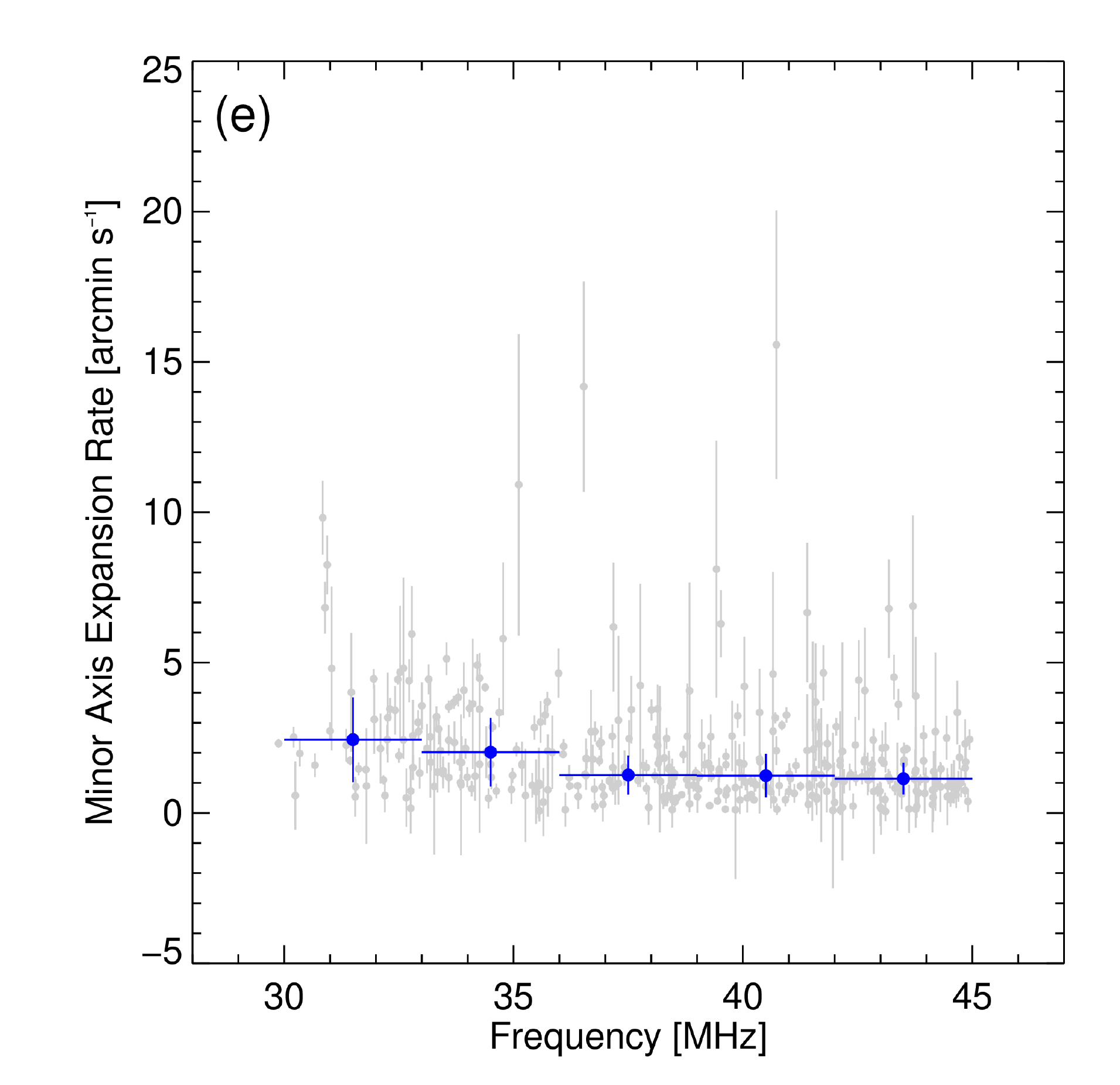}
    \endminipage\hfill
    \minipage{0.32\textwidth}%
      \includegraphics[width=\linewidth]{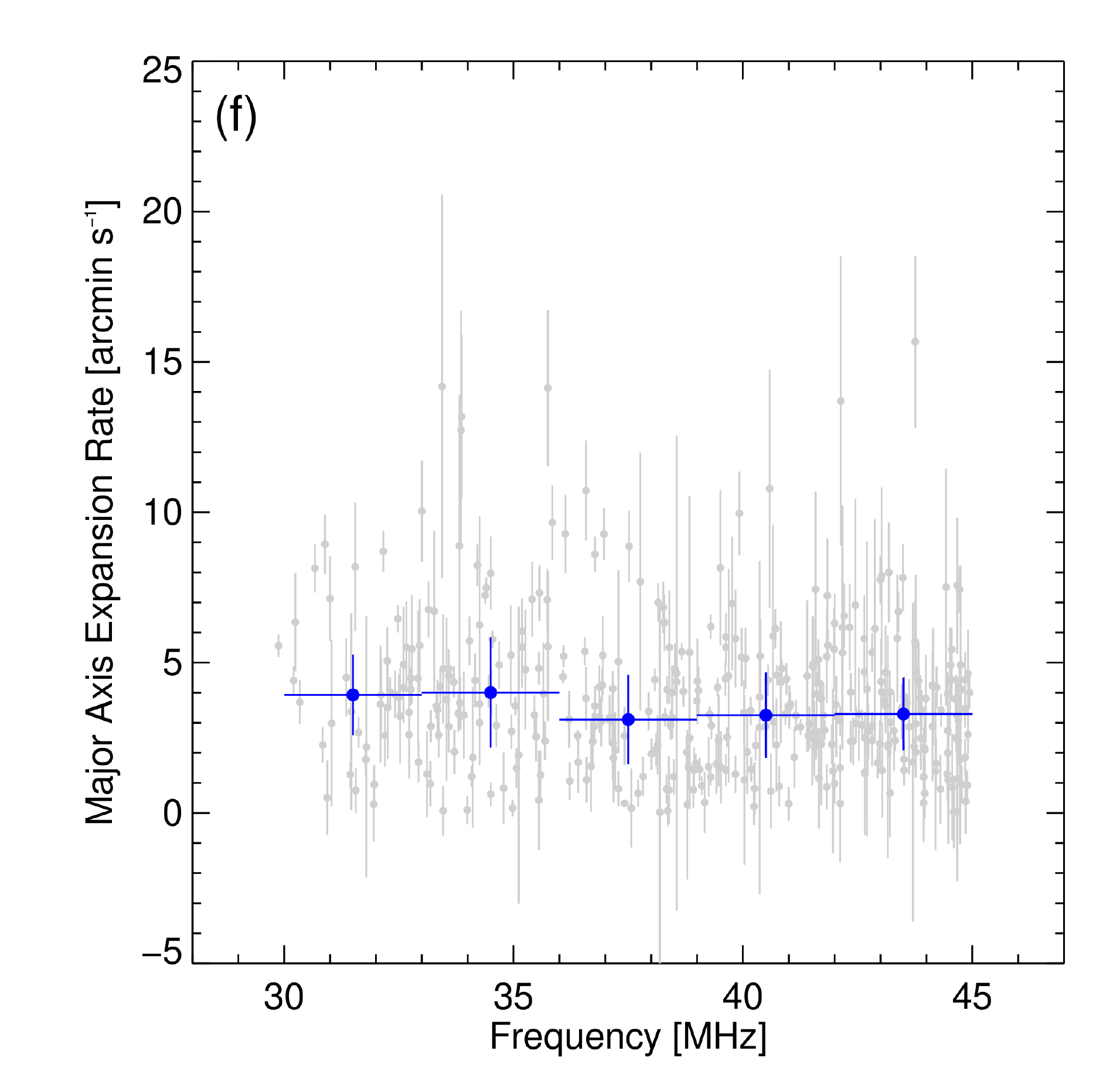}
    \endminipage
    \caption{Spike characteristics derived from imaging observations. The grey data show the observed quantity and associated uncertainties. The blue points show the median values across 3 MHz bins, with the vertical error as the interquartile range between the 25th and 75th percentiles, and a power-law fit given by the solid blue line. \textbf{(a)} Observed FWHM area. \textbf{(b)} FWHM minor axis size. \textbf{(c)} FWHM major axis size. \textbf{(d)} Areal expansion rate. \textbf{(e)} Minor axis expansion rate. \textbf{(f)} Major axis expansion rate.}
    \label{fig:imaging_char}
\end{figure*}

\begin{figure*}[!htb]
    \minipage{0.32\textwidth}
      \includegraphics[width=\linewidth]{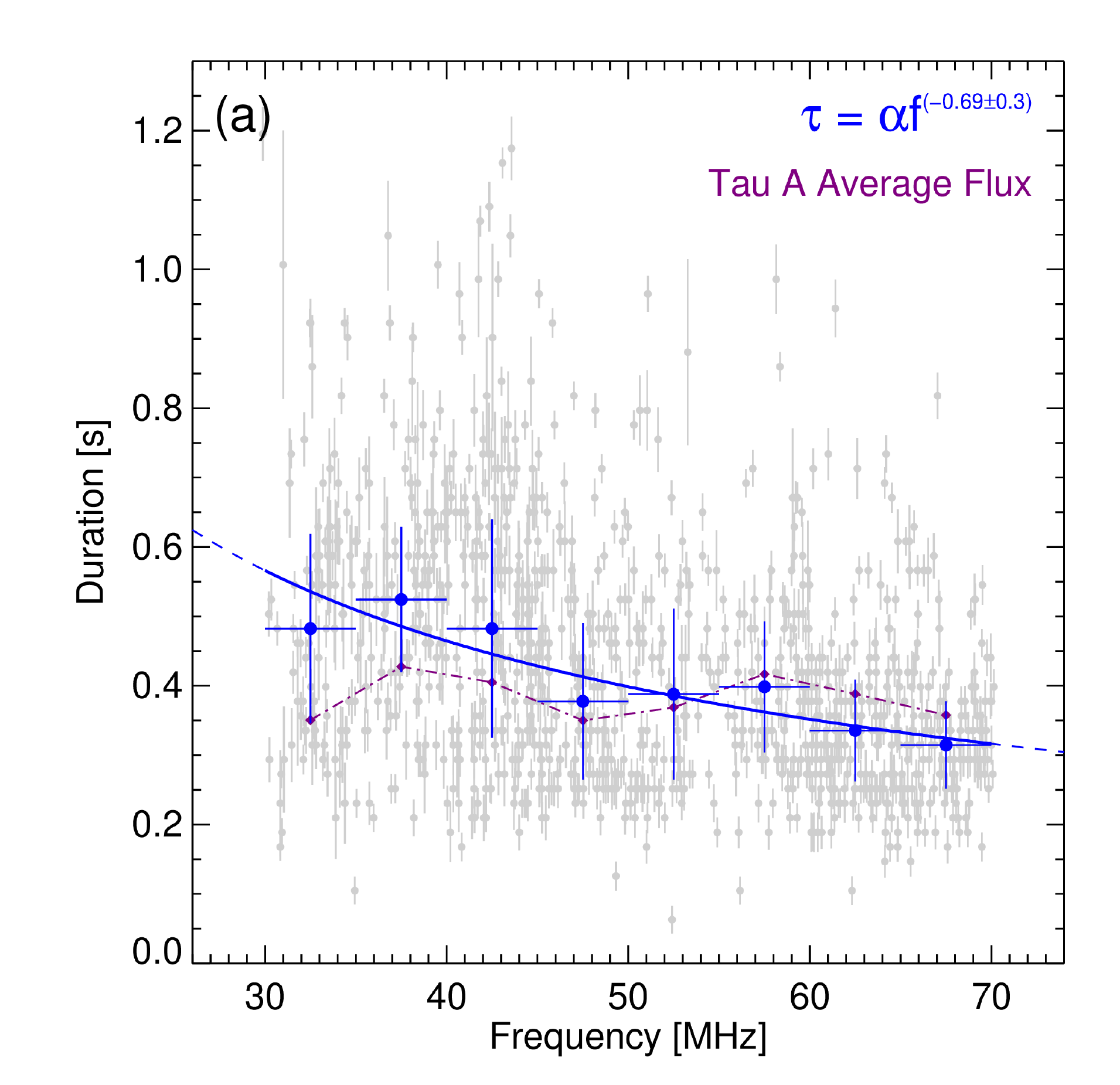}
    \endminipage\hfill
    \minipage{0.32\textwidth}
      \includegraphics[width=\linewidth]{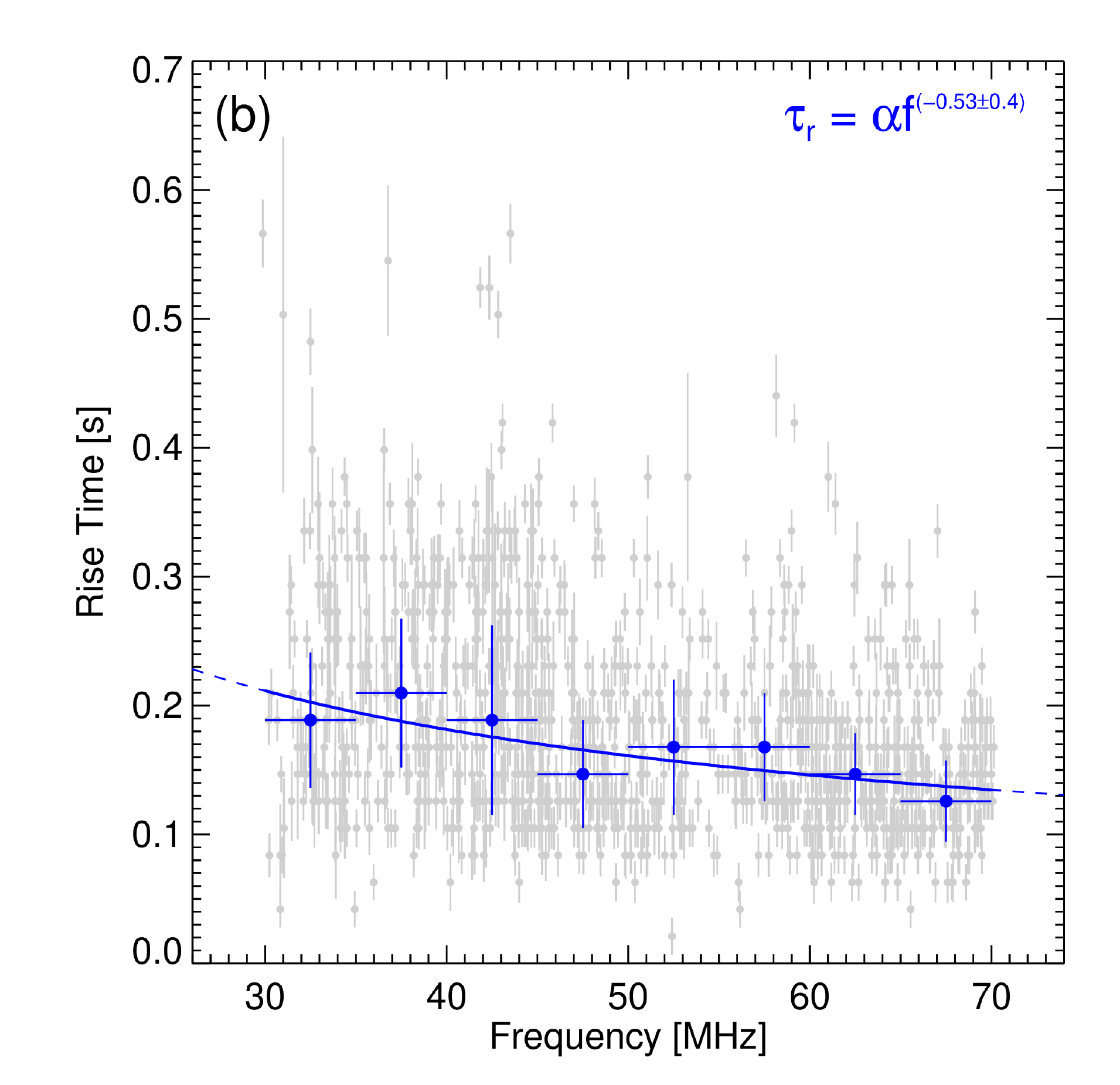}
    \endminipage\hfill
    \minipage{0.32\textwidth}%
      \includegraphics[width=\linewidth]{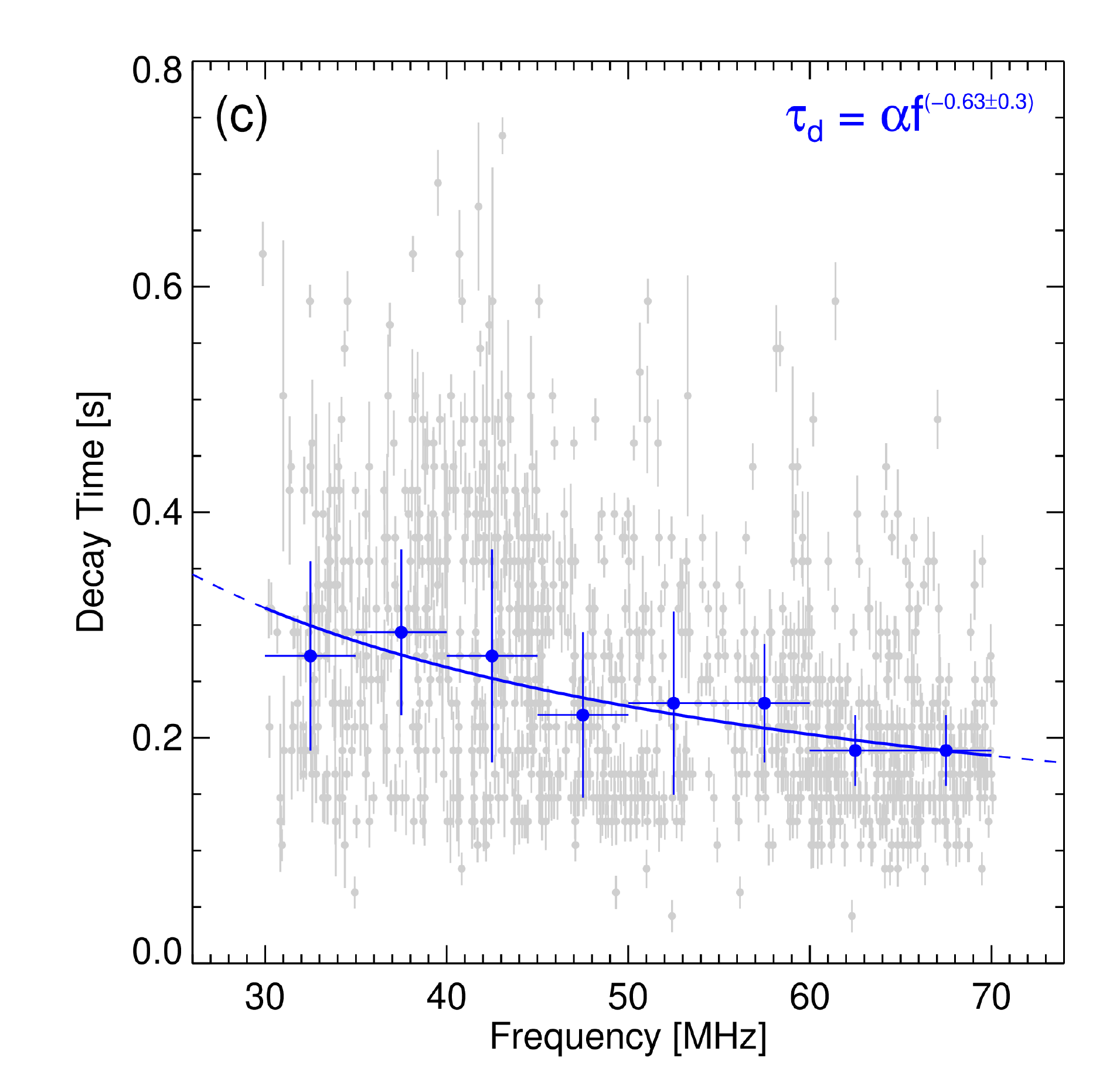}
    \endminipage\hfill
    \minipage{0.32\textwidth}
      \includegraphics[width=\linewidth]{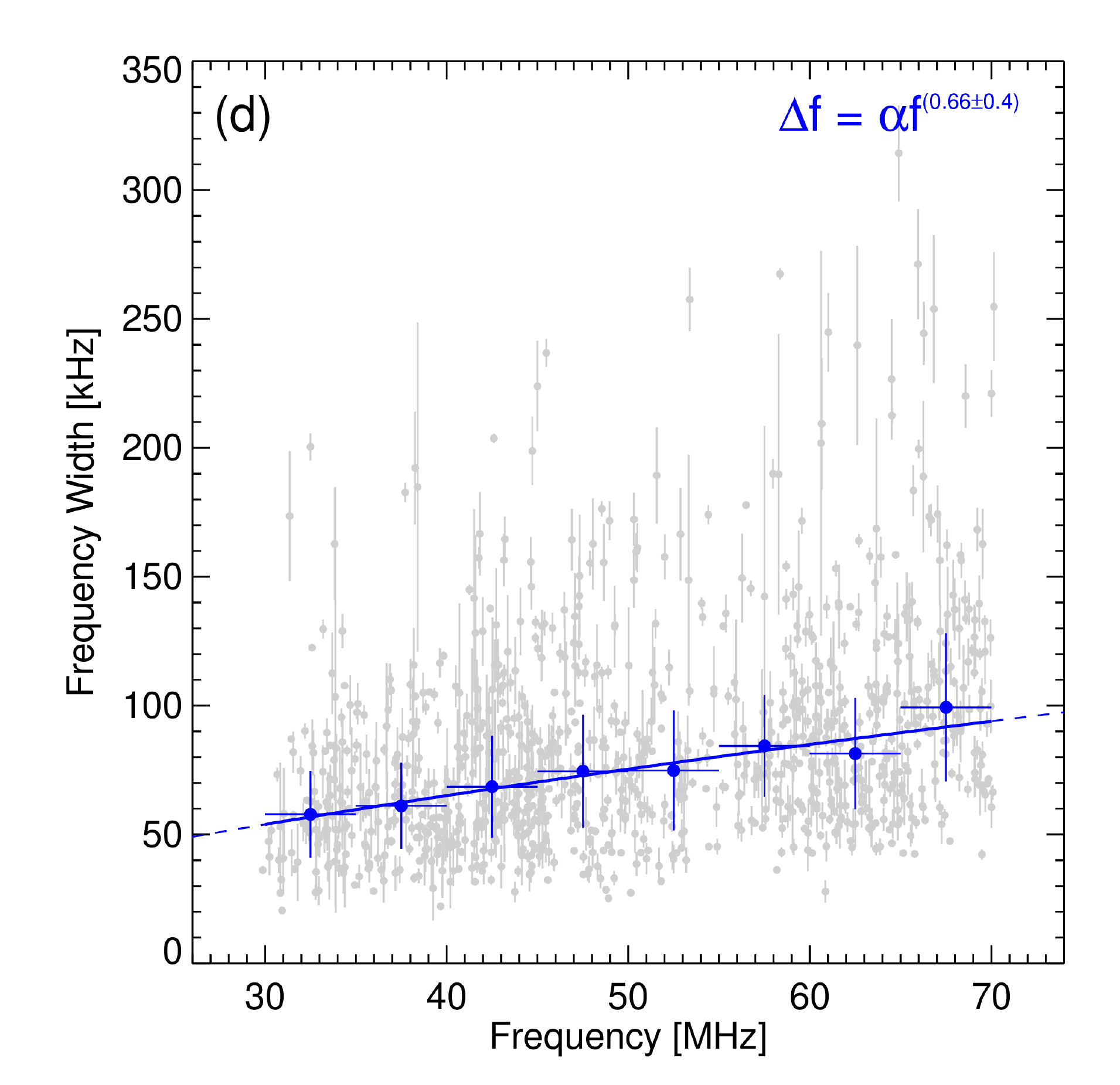}
    \endminipage\hfill
    \minipage{0.32\textwidth}
      \includegraphics[width=\linewidth]{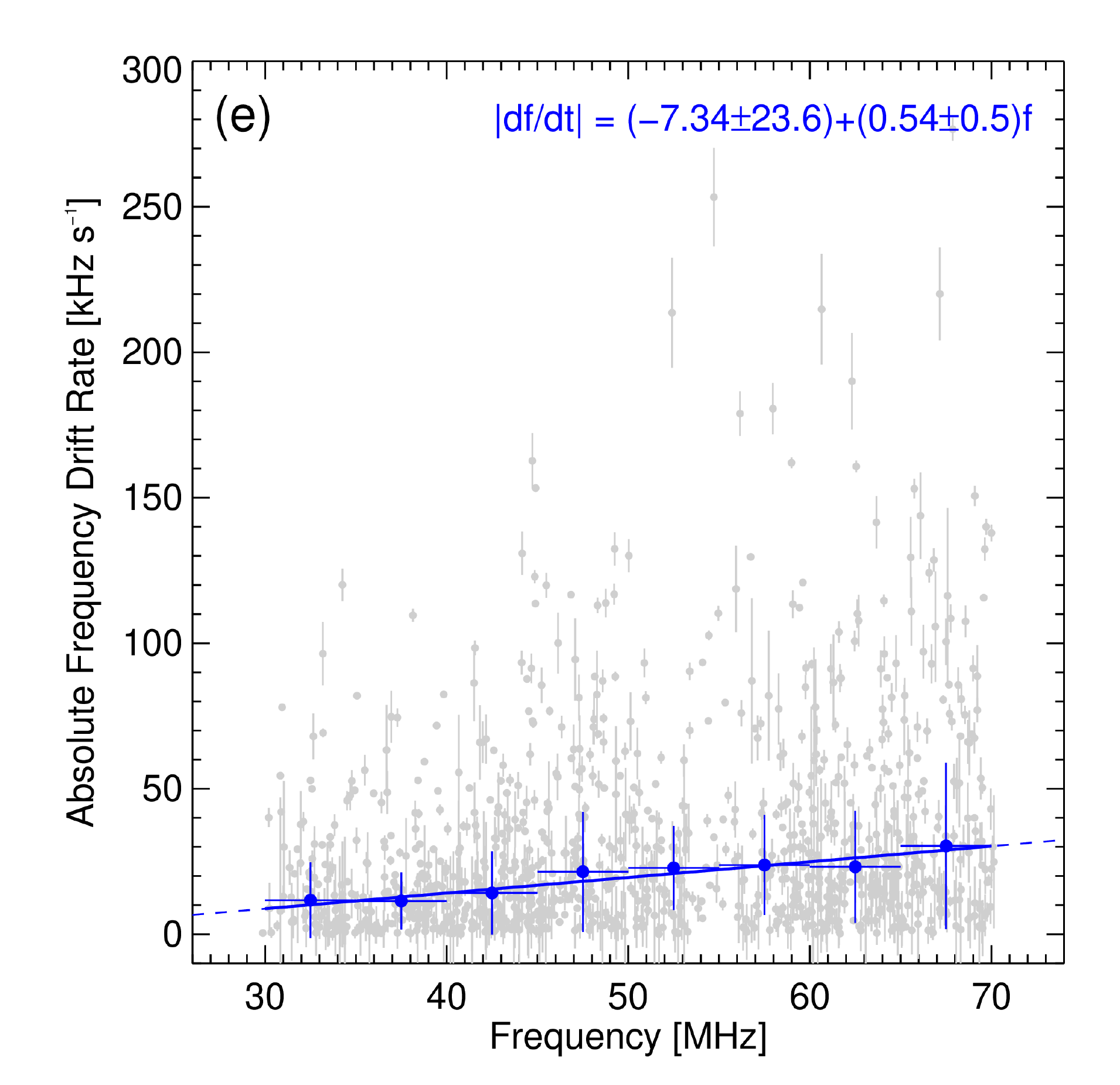}
    \endminipage\hfill
    \minipage{0.32\textwidth}
      \includegraphics[width=\linewidth]{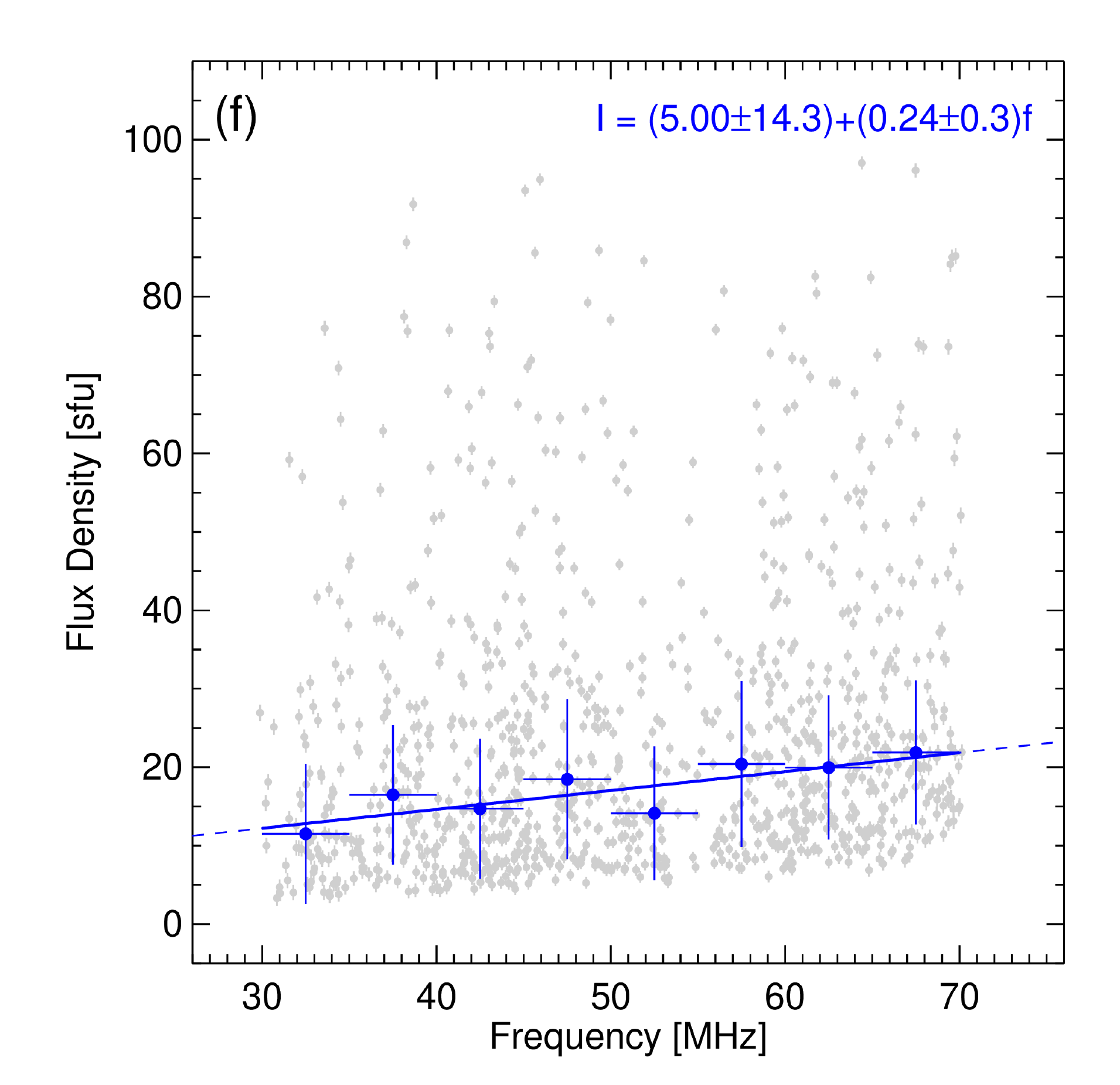}
    \endminipage
    \caption{Spike characteristics derived from dynamic spectra observations. The light gray points show the data with associated uncertainties. The blue points show the median values across 5 MHz bins, with the vertical error as the interquartile range representing the 25th and 75th percentiles, and a power-law or linear fit shown by the blue line. \textbf{(a)} FWHM Duration. The purple points show the median Tau A flux on the day of observation. \textbf{(b)} Rise Time at the half-maximum intensity level. \textbf{(c)} Decay time at the half-maximum intensity level. \textbf{(d)} FWHM bandwidth. \textbf{(e)} Absolute frequency drift rate. \textbf{(f)} Flux at the lightcurve peak.}
    \label{fig:ds_char}
\end{figure*}

\end{document}